\documentclass{birkjour}
 \newtheorem{thm}{Theorem}[section]

 \theoremstyle{definition}
 
 \theoremstyle{remark}
 \newtheorem{rem}[thm]{Remark}
 
 \numberwithin{equation}{section}

\usepackage{url}
\usepackage{amsmath,amssymb}
\usepackage{amsfonts}
\usepackage{subfig} % for subfigures
\usepackage{caption}
\begin{document}
%--------------------------------------%
\title[Visualization and machine learning for forecasting of COVID-19 in Senegal]
 {Visualization and machine learning for forecasting of COVID-19 in Senegal }
 
%----------Author 1
\author[Ndiaye et al.]{Babacar Mbaye Ndiaye, Mouhamadou A.M.T. Balde, Diaraf Seck$^*$}
\address{Laboratory of Mathematics of Decision and Numerical Analysis.\br
University of Cheikh Anta Diop. 
BP 45087, 10700. Dakar, Senegal.\br
$^*$IRD, UMMISCO, Dakar, Senegal.}

\email{babacarm.ndiaye@ucad.edu.sn, mouhamadouamt.balde@ucad.edu.sn,\\ diaraf.seck@ucad.edu.sn}

\thanks{This work was completed with the support of the NLAGA project}


\date{May 16, 2020}
% ----------------------------------------------------------------------%
\begin{abstract} 
In this article, we give visualization and different machine learning technics for two weeks and 40 days ahead forecast based on public data.
On July 15, 2020, Senegal reopened its airspace doors, while the number of confirmed cases is still increasing. The population no longer respects hygiene measures, social distancing as at the beginning of the contamination. Negligence or tiredness to always wear the masks? We make forecasting on the inflection point and possible ending time.
\end{abstract}
%
% ------------------------------------------------------------------%
\maketitle
% ------------------------------------------------------------------%
 \vspace{-1.25cm}
\section{Introduction}
\noindent  In Senegal, the pandemic of COVID-19 was reported in Dakar on March 2th, 2020, and then quickly spread out national wide. The pandemic is in its active phase and is progressing at breakneck speed. The latest assessment, after more than 147 days (by July 26, 2020) of struggle, showed a worrying health situation with an increase in the number of confirmed cases.
Current social distancing measures to impede COVID-19 are economically unsustainable in the long term. Models are needed to understand the implications of possible relaxation options for these measures. 
In \cite{ndiaye1,ndiaye2,ndiaye3,balde}, authors use deterministic, stochastic models and machine learning technics to study and forecast the COVID-19 cases in Senegal. In these works, we take into account the nationwide measures in Senegal.
In \cite{sarr}, we explore some contamination factors on the COVID-19 evolution in Senegal (population density, religions/beliefs, food, youth, temperature, humidity, and cross-immunity) that reduce the rate of spread.\\
\noindent As of July 2, 2020, Senegal had just completed four months of response to the COVID-19
disease. Epidemiological surveillance had remained static for a long time with the investigation of suspected and confirmed cases and monitoring cases of contact.\\
\noindent In this work, first, we collect the pandemic data carefully from \cite{msas, datahub}, from  2020, March 02, to July 26. Second, we propose visualization and different machine learning
technics (linear regression, polynomial regression, support vector regression, prophet, and multilayer perceptron) to analyze the coronavirus pandemic in Senegal.\\
\noindent We organize the article as follows. In section \ref{analysis}, we present some data analysis followed by the visualization in section \ref{viz}. In section \ref{prediction}, we perform several machine learning technics for forecasting for two weeks and 40 days. Finally, in section \ref{ccl}, we present conclusions and perspectives.
\vspace{-.5cm}

%-----%
\section{Data analysis}\label{analysis}
\begin{figure}[h!]
  \subfloat[Senegal COVID-19 cases - confirmed, deaths and recovered]{
	\begin{minipage}[1\width]{0.5\textwidth}
	   \centering
	   \includegraphics[width=1.1\textwidth]{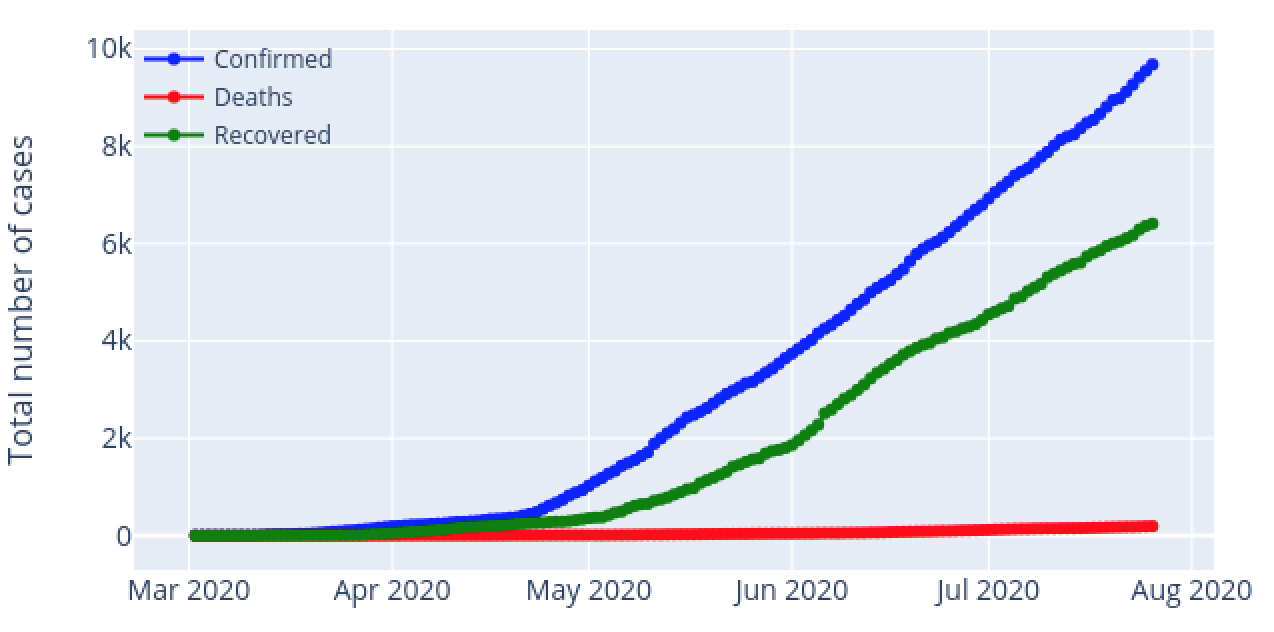}\label{crd_sn}
	\end{minipage}}
% \hfill 	
  \subfloat[Frequency histogram of confirmed cases]{
	\begin{minipage}[1\width]{ 0.5\textwidth}
	   \centering
	   \includegraphics[width=1.05\textwidth]{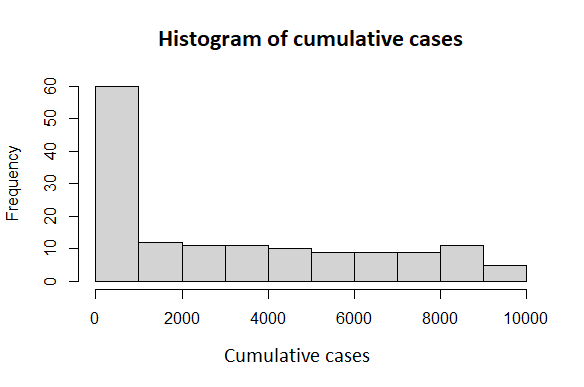}\label{Histoc}
	\end{minipage}}
\caption{Cumulative cases}\label{sn_crd-histo}
\end{figure}

The simulations are carried out from data in \cite{msas, datahub}, from 2020, March 02 to July 26. The numerical tests were performed by using the Python with Panda library \cite{python} and nnfor R package \cite{CRAN}, on a computer with the following characteristics:
intel(R) Core-i7 CPU 2.60GHz, 24.0Gb of RAM, under UNIX system. According to daily
reports, we first analyze and make some data preprocessing before simulations. Figure \ref{crd_sn} illustrates the cumulative numbers of confirmed, recovered, and death cases, and Figure \ref{Histoc} is a frequency histogram which analyzes how many times values of an interval of cumulative confirmed cases are reached. Then, we get various summary statistics (per day), by giving the mean, standard deviation, minimum and maximum values, and the quantiles of the data (see Tables \ref{stat_confcases_1} and \ref{stat_confcases_2}).
\vspace{-.5cm}

%table
\begin{table}[h!] 
\begin{center}
\begin{tabular}{|c|c|c|c|c|c|c|} 
 \hline
 {\bf values} 	& {\bf tests} & 	{\bf cases} & {\bf	contact} &{\bf	imported}& 	{\bf community}& 	{\bf confirmed} \\ \hline
mean 	& 712.75	& 65.85	 &54.52	&1.19	&10.18	 &3114.05	 \\ 
\hline
std &	492.20 	  &49.06	  &41.74	&3.00	&10.93	&3104.67	  \\ 
\hline
min 	&  1   &  0 &  	0 &     0 	 &    0   &   	1 \\ 
\hline
25\% 	& 158.00 &	12.00&	8.00	&0.00	 & 1.00  & 	240.50  \\ 
\hline
50\% &	827.00	&71.00	& 63.00&	0.00&	7.00 &	2189.00 \\ 
\hline
75\% &	1087.75 &	105.50	& 87.50&	1.00&	15.00 &	5711.00\\ 
\hline
max &1820 &	177	&169	&27&	49&  9681\\
\hline
\end{tabular}
\end{center}
\caption{Senegal summary statistics (per day) until July 26th, 2020 (tests, cases, 	contact, imported,  community and confirmed).}\label{stat_confcases_1}
\end{table}
%table
\begin{table}[h!] 
\begin{center}
\begin{tabular}{|c|c|c|c|c|c|c|} 
 \hline
 {\bf values} 	& {\bf recovered} &{\bf deaths} 	& {\bf evacuated} 	& {\bf severe} &	{\bf active} &	{\bf ratios} \\ \hline
mean 	& 1907.19	& 47.61&	0.79	& 13.27 & 	1159.24  &	0.13 \\ 
\hline
std &  2116.01 	& 56.99	& 0.41 	& 15.01	  & 974.17   &	0.18 	 \\ 
\hline
min 	&  0     &  0    &  0   &    0  &      1   &     0    \\ 
\hline
25\% 	& 109.00	& 2.00 &	1.00&	0.00&	131.00	  & 0.07	 \\ 
\hline
50\% & 842.00 &	23.00  &	1.00 	 & 8.00   &	1324.00	& 0.09  \\ 
\hline
75\% &	3823.50 &	80.50 & 1.00 &22.00 &	1840.50	&0.11  \\ 
\hline
max &  6409 &	191 &	1	& 53 &	3081   & 1\\
\hline
\end{tabular}
\end{center}
\caption{Senegal summary statistics (per day) until July 26th, 2020 (recovered, deaths, evacuated, severe, active and ratios).}\label{stat_confcases_2}
\end{table}
\vspace{-.5cm}

\begin{figure}[h!]
  \subfloat[number of tests per day]{
	\begin{minipage}[1\width]{0.5\textwidth}
	   \centering
	   \includegraphics[width=1\textwidth]{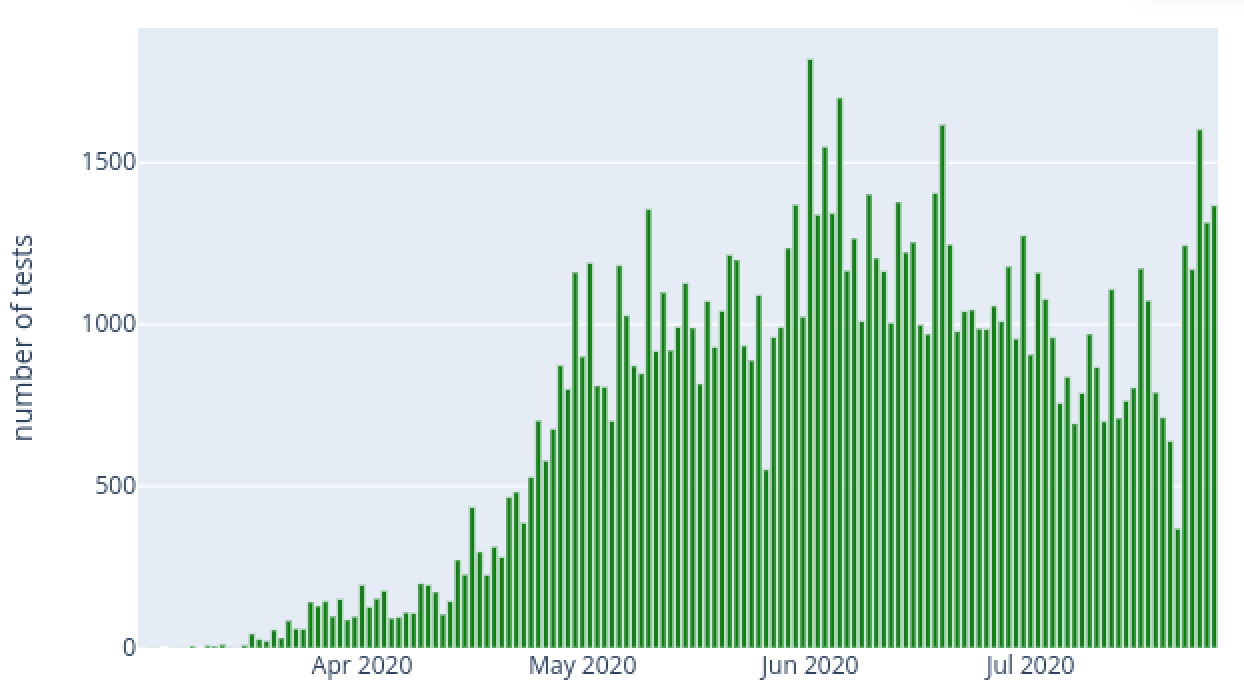}\label{tests}
	\end{minipage}}
% \hfill 	
  \subfloat[ratios per day]{
	\begin{minipage}[1\width]{ 0.5\textwidth}
	   \centering
	   \includegraphics[width=1\textwidth]{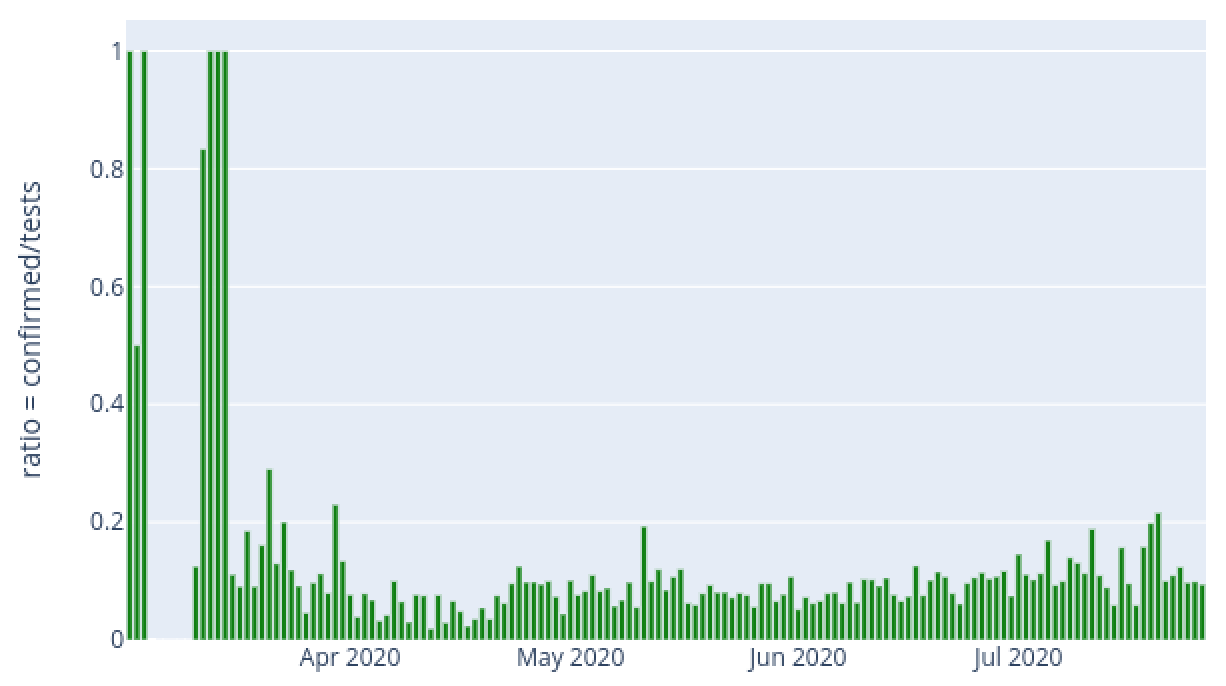}\label{ratios}
	\end{minipage}}
\newline
  \subfloat[severe cases per day]{
	\begin{minipage}[1\width]{ 0.5\textwidth}
	   \centering
	   \includegraphics[width=1\textwidth]{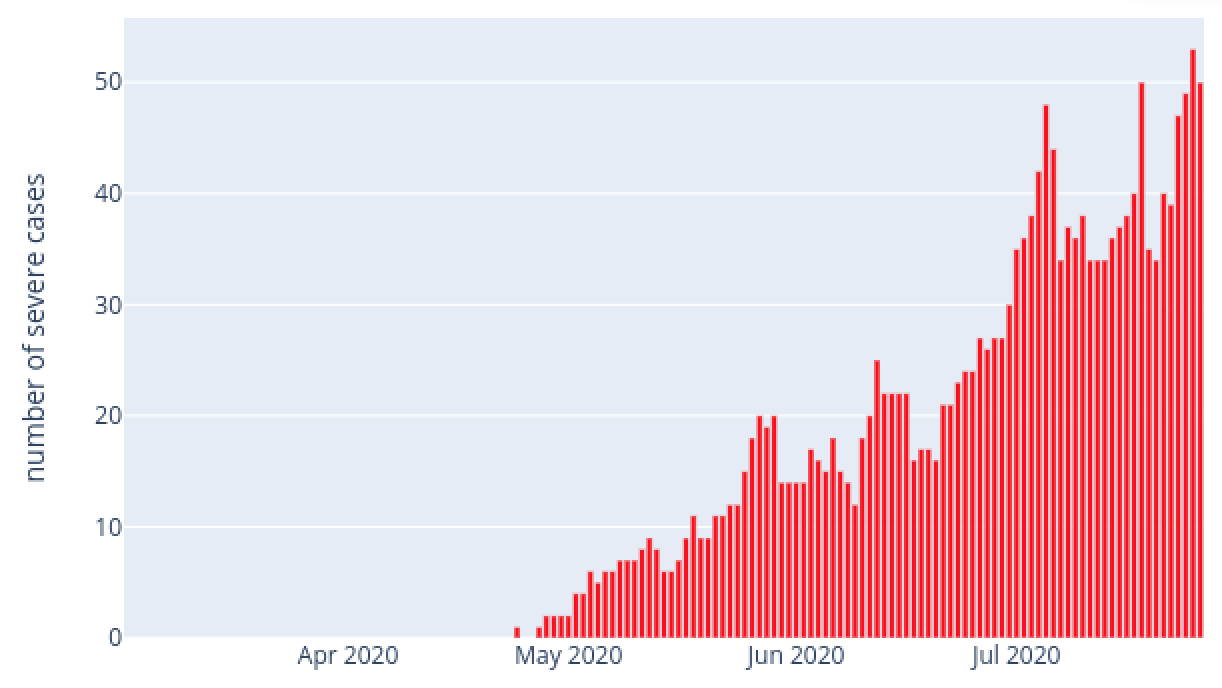}\label{severe}
	\end{minipage}}
% \hfill 	
  \subfloat[community cases per day]{
	\begin{minipage}[1\width]{ 0.5\textwidth}
	   \centering
	   \includegraphics[width=1\textwidth]{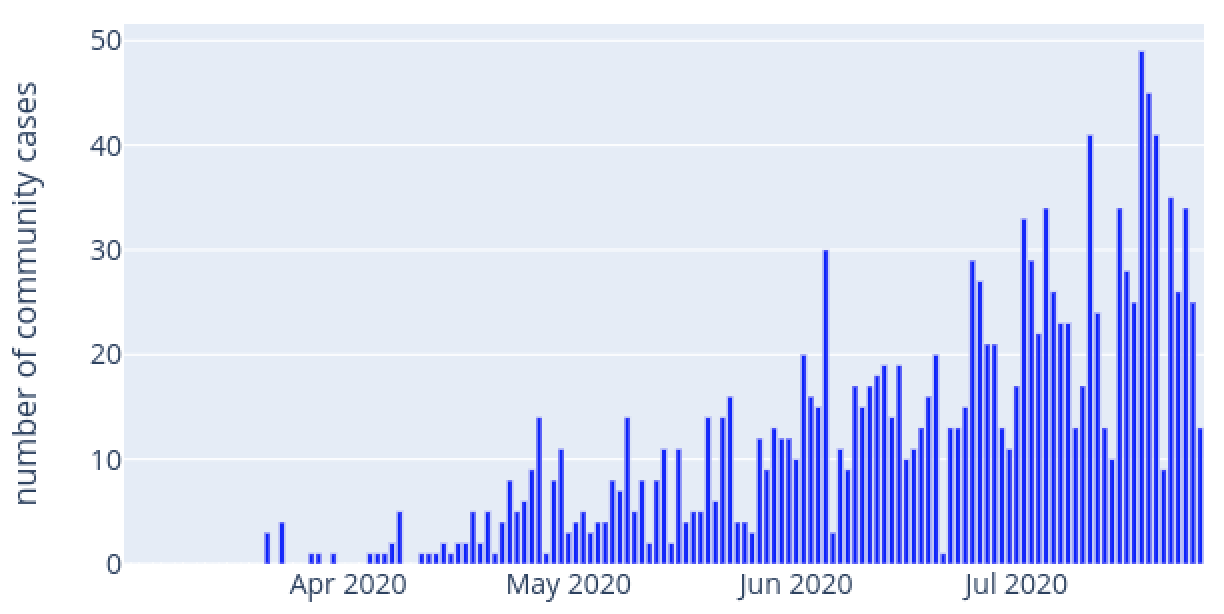}\label{comty}
	\end{minipage}}
\caption{Senegal: tests - ratios / community and severe cases}\label{testsresults}
\end{figure}

\noindent Let's define the Active cases = Confirmed-Recovered-Deaths, the  Closed cases = Recovered + Deaths. 
The number of performed tests par day is not a constant (see Figure \ref{tests}). The Figure \ref{ratios} shows that in Senegal, between March 02 and July 26, 2020, the ratio  (ratio=confirmed cases/tests) is almost constant despite variations in the number of daily tests performed and considered to be unrepresentative. Despite official statements about the peak period, it seems difficult to say whether the peak of contamination has already been reached or will even be reached shortly. 
In addition, we find that the ratio varies on average by 6 to 10\%. 
Figures \ref{severe} and \ref{comty} show that the maximum number of severe and community cases are less than 1\% (0.54\% and 0.5\%, respectively) of the number of confirmed cases.

%---%
\section{Visualization}\label{viz}
\subsection{Datewise analysis}
\begin{figure}[h!]
  \subfloat[distribution of active cases]{
	\begin{minipage}[1\width]{0.5\textwidth}
	   \centering
	   \includegraphics[width=1\textwidth]{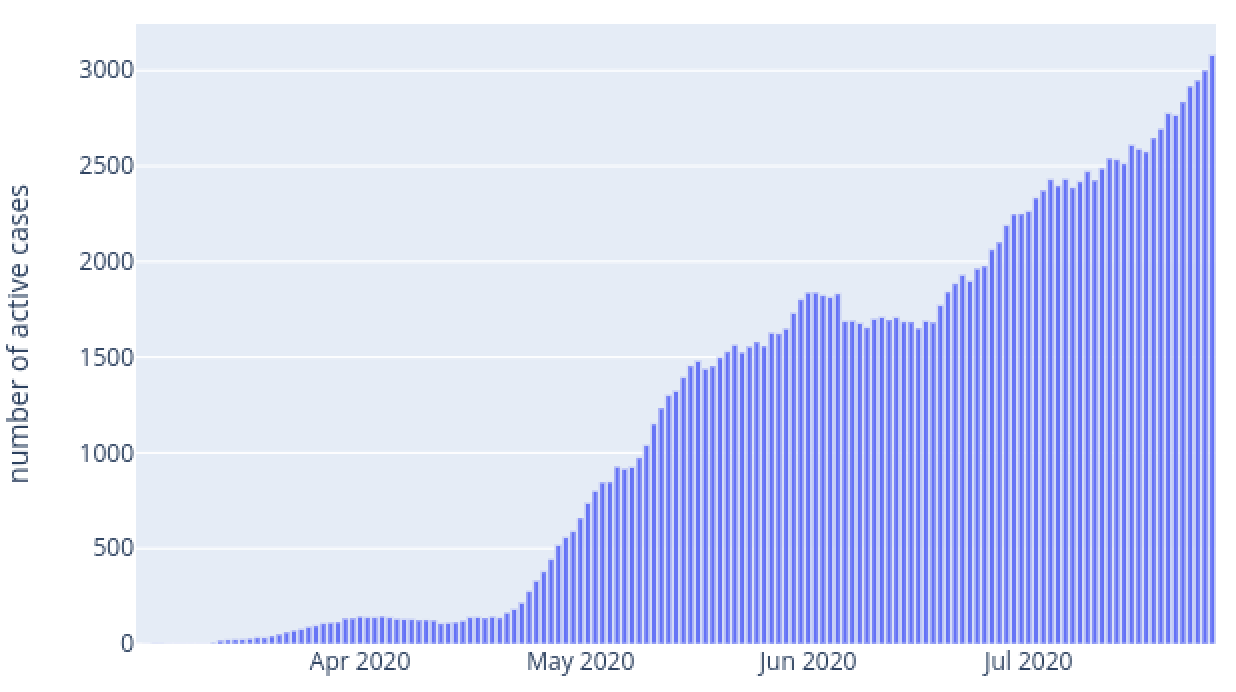}\label{sn_acticases}
	\end{minipage}}
% \hfill 	
  \subfloat[distribution of closed cases]{
	\begin{minipage}[1\width]{ 0.5\textwidth}
	   \centering
	   \includegraphics[width=1\textwidth]{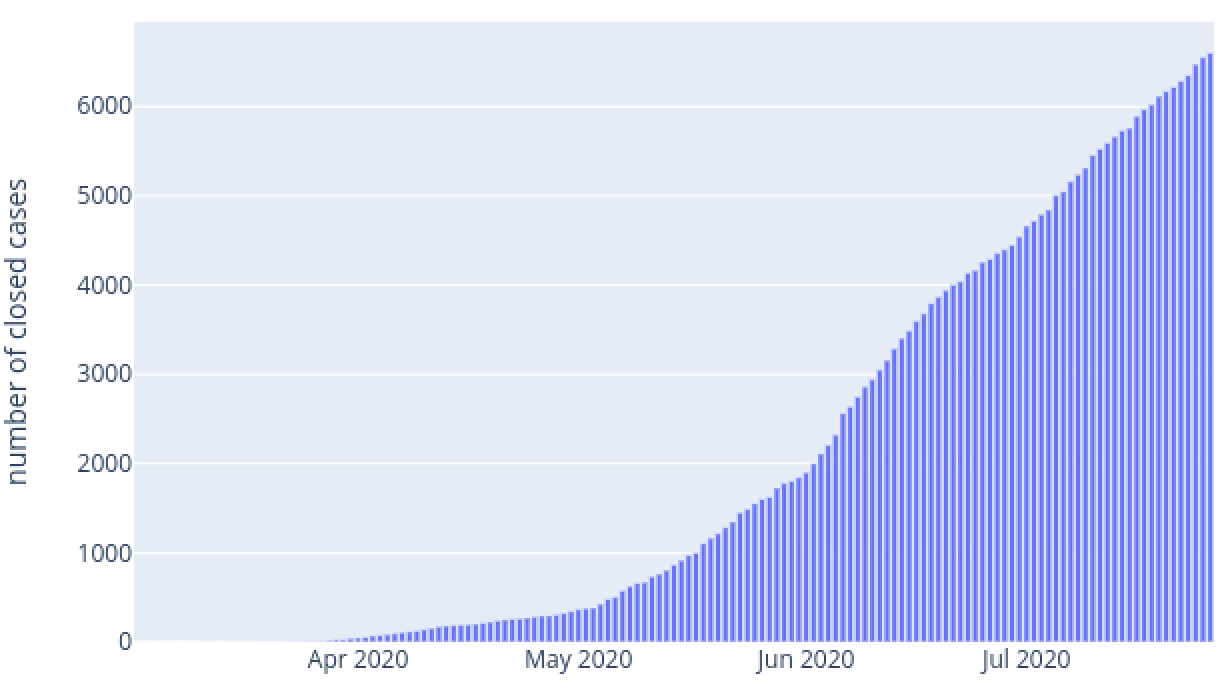}\label{sn_closcases}
	\end{minipage}}
\caption{Number of active and closed cases per day}\label{sn_activcloscases}
\end{figure}

\begin{figure}[h!]
	\centering
	\includegraphics[width=.8\linewidth]{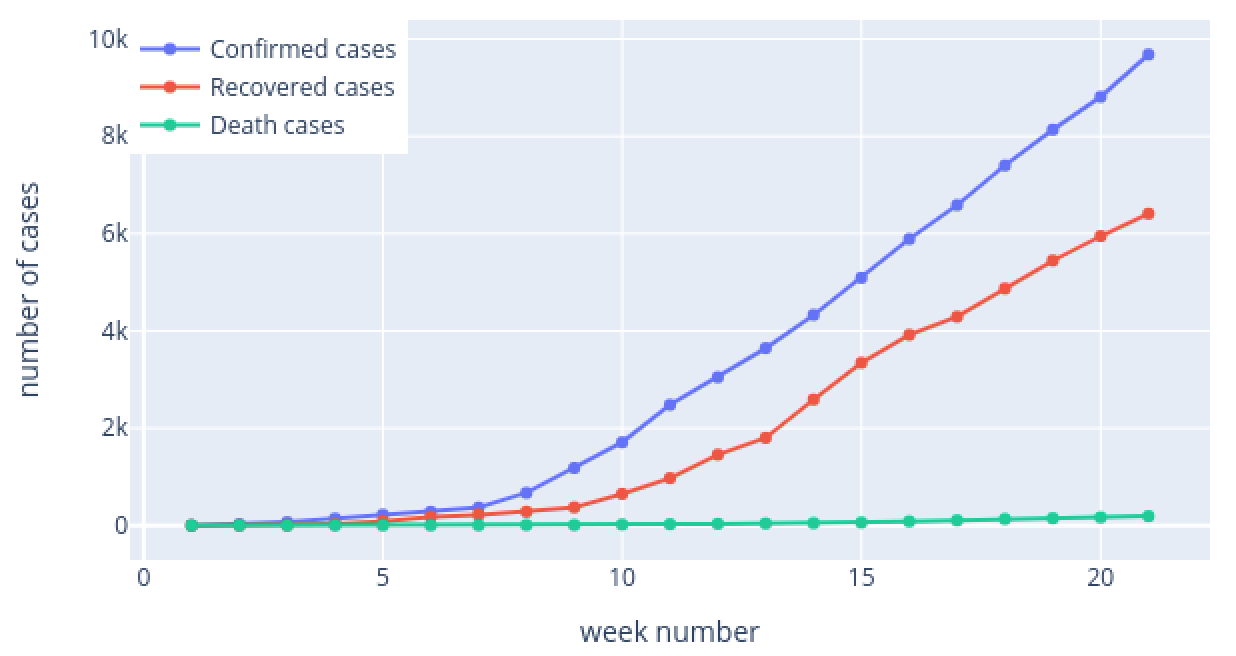}
		\par\vspace{-0.25cm}
	\caption{Weekly growth of confirmed, recovered and death cases}\label{sn_weeklygrowth}
\end{figure}

\begin{figure}[h!]
	\centering
	\includegraphics[width=.95\linewidth]{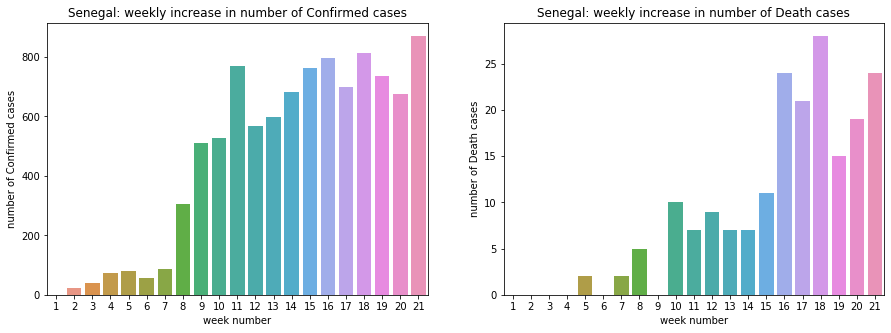}
	\caption{Weekly increase in number of confirmed and death cases}\label{sn_weeklyincrease}
\end{figure}

We have: Active =  Confirmed - Recovered - Deaths, and Closed = Recovered + Deaths. The grouping of different types of cases as per the date is given by the Table \ref {basic_info}.
\begin{table}[h!] 
\begin{center}
\begin{tabular}{|l|c|} 
 \hline
 {\bf basic information} 	& {\bf value}  \\ \hline
total number of confirmed cases &  9681\\\hline
total number of recovered cases  & 6409\\\hline
total number of deaths cases  &  191\\\hline
total number of active cases &  3081\\\hline
total number of closed cases &  6600\\\hline
\hline
approximate number of confirmed cases per day &  66.0\\\hline
approximate number of recovered cases per day    &   44.0\\\hline
approximate number of death cases per day   &  1.0\\\hline
\hline
approximate number of confirmed cases per hour  &  3\\\hline
approximate number of recovered cases per hour  &  2\\\hline
approximate number of death cases per hour  &  0\\\hline 
\hline
number of confirmed cases in last 24 hours   &   129\\\hline
number of recovered cases in last 24 hours   &  45\\\hline
number of death cases in last 24 hours   &  4\\\hline 
\end{tabular}
\end{center}
\caption{Senegal: disease spread, by July 26th 2020.}\label{basic_info}
\end{table}
\noindent The number of actives and closed cases are given in Figures \ref{sn_acticases} and \ref{sn_closcases}.
\noindent We see that the number of closed cases is increasing, which is positive for Senegal.
\noindent The weekly growth of confirmed, recovered and death cases is given by Figure \ref{sn_weeklygrowth}, while the weekly increase in number of confirmed and death cases is given by Figure  \ref{sn_weeklyincrease}.
\noindent We see that weeks 16, 18 and 21 were fatals for deaths and confirmed cases.  \\

\subsection{High population density}\label{popdensity}
\begin{figure}[h!]
  \subfloat[Dakar north]{
	\begin{minipage}[1\width]{0.48\textwidth}
	   \centering
	   \includegraphics[width=1.\textwidth]{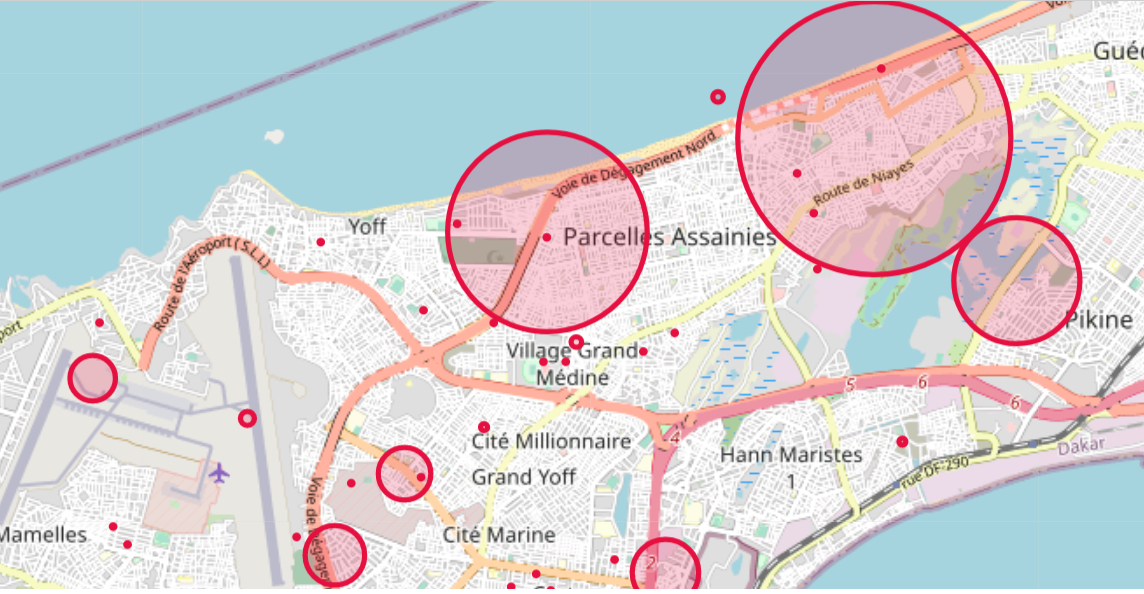}\label{dkrGPM}
	\end{minipage}}
% \hfill 	
  \subfloat[Dakar west and Dakar center]{
	\begin{minipage}[1\width]{ 0.48\textwidth}
	   \centering
	   \includegraphics[width=1.\textwidth]{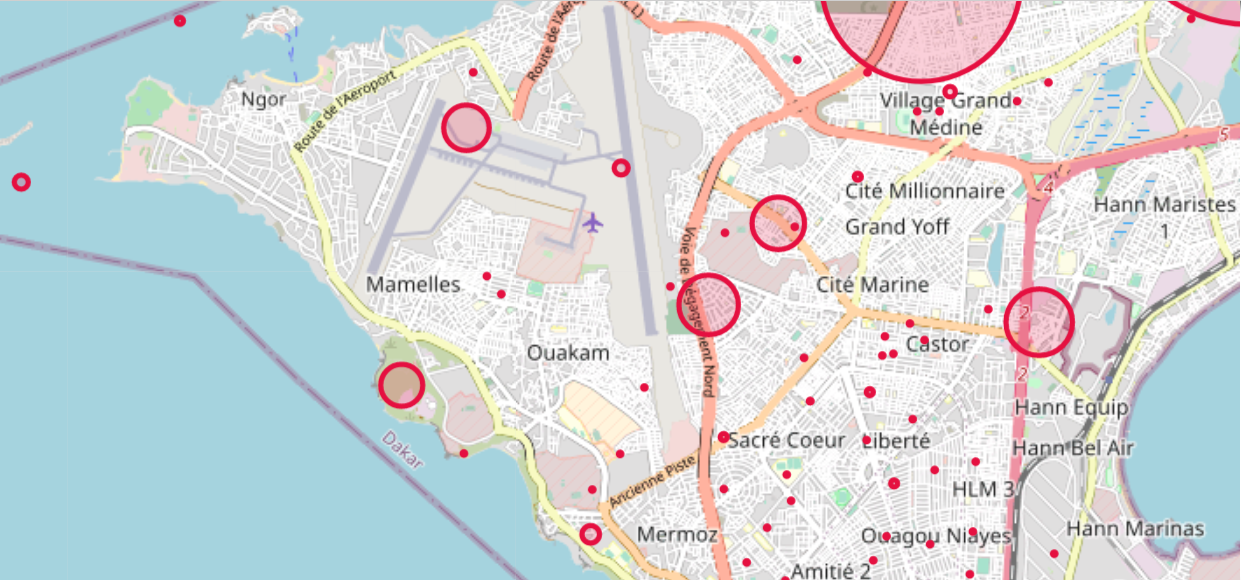}\label{dkr_center}
	\end{minipage}}
\newline
  \subfloat[Dakar south]{
	\begin{minipage}[1\width]{ 0.48\textwidth}
	   \centering
	   \includegraphics[width=1.\textwidth]{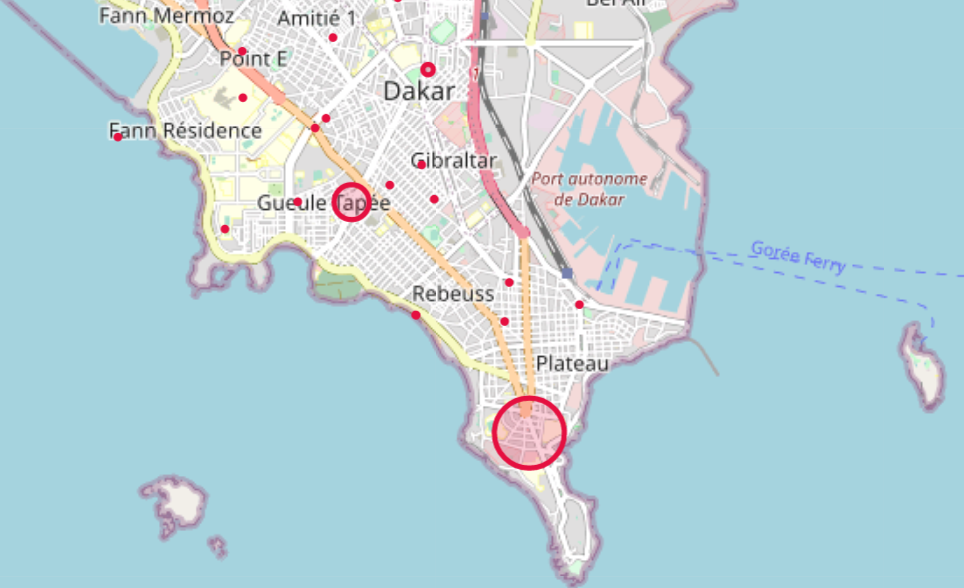}\label{dkr_sud}
	\end{minipage}}
%
% \hfill 	
  \subfloat[zoom of Dakar cases]{
	\begin{minipage}[1\width]{ 0.48\textwidth}
	   \centering
	   \includegraphics[width=1.\textwidth]{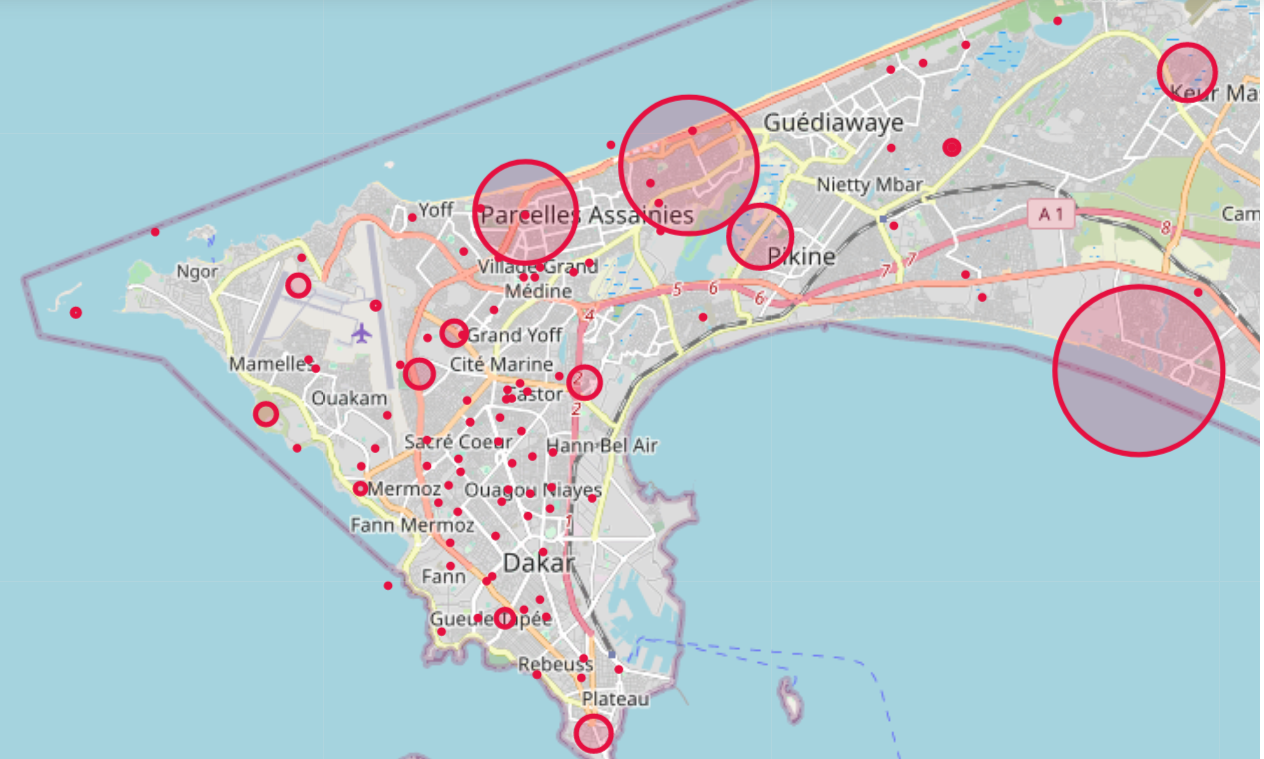}\label{dkr_zoom}
	\end{minipage}}
\caption{Dakar region cases}\label{dakar_viz}
\end{figure}

\begin{figure}[h!]
  \subfloat[the first 5 zones]{
	\begin{minipage}[1\width]{0.49\textwidth}
	   \centering
	   \includegraphics[width=1.\textwidth]{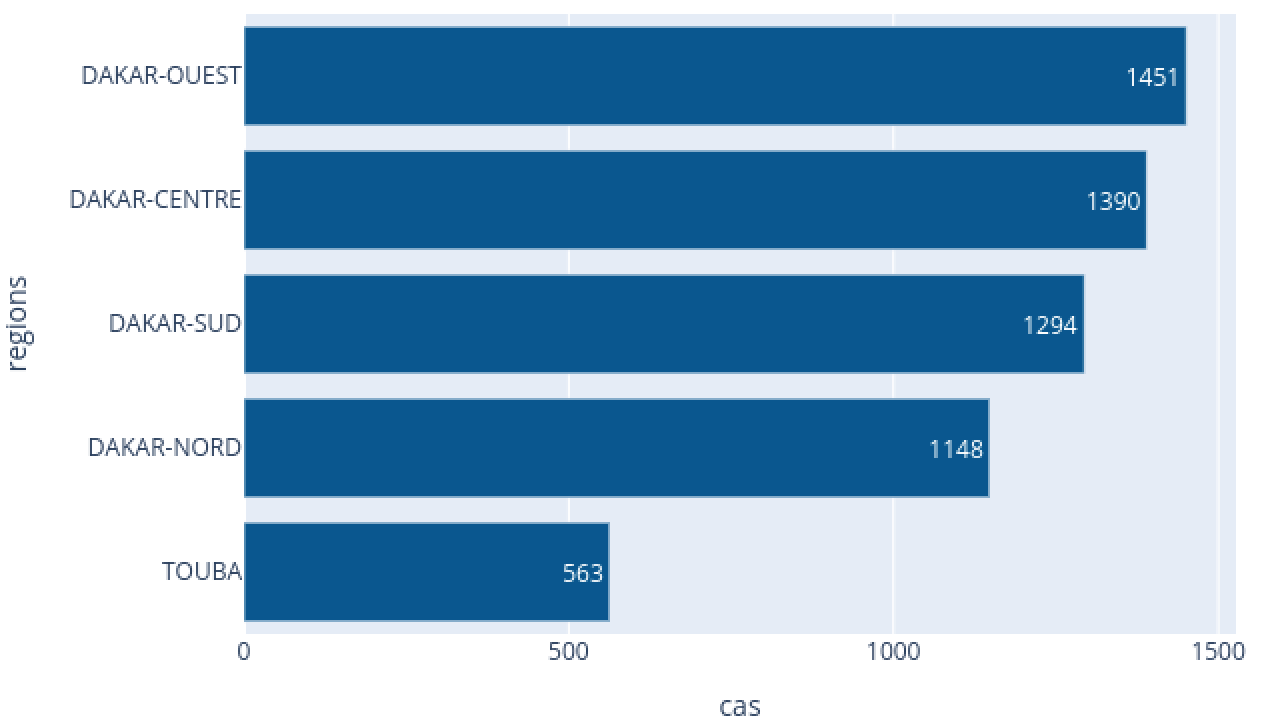}\label{first5_zones}
	\end{minipage}}
% \hfill 	
  \subfloat[the first 10 zones]{
	\begin{minipage}[1\width]{ 0.49\textwidth}
	   \centering
	   \includegraphics[width=1.\textwidth]{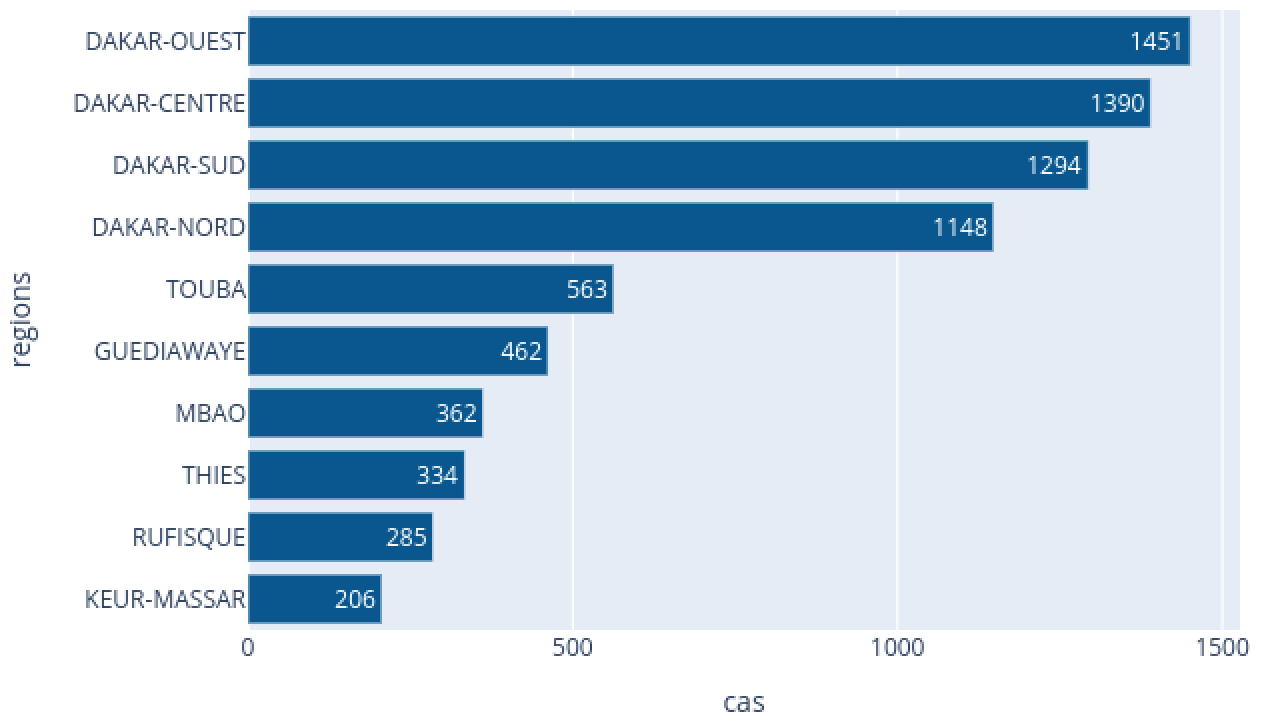}\label{first10_zones}
	\end{minipage}}
\caption{Number of confirmed cases per zone}\label{caseszones}
\end{figure}

\begin{figure}[h!]
	\centering
	\includegraphics[width=0.8\linewidth]{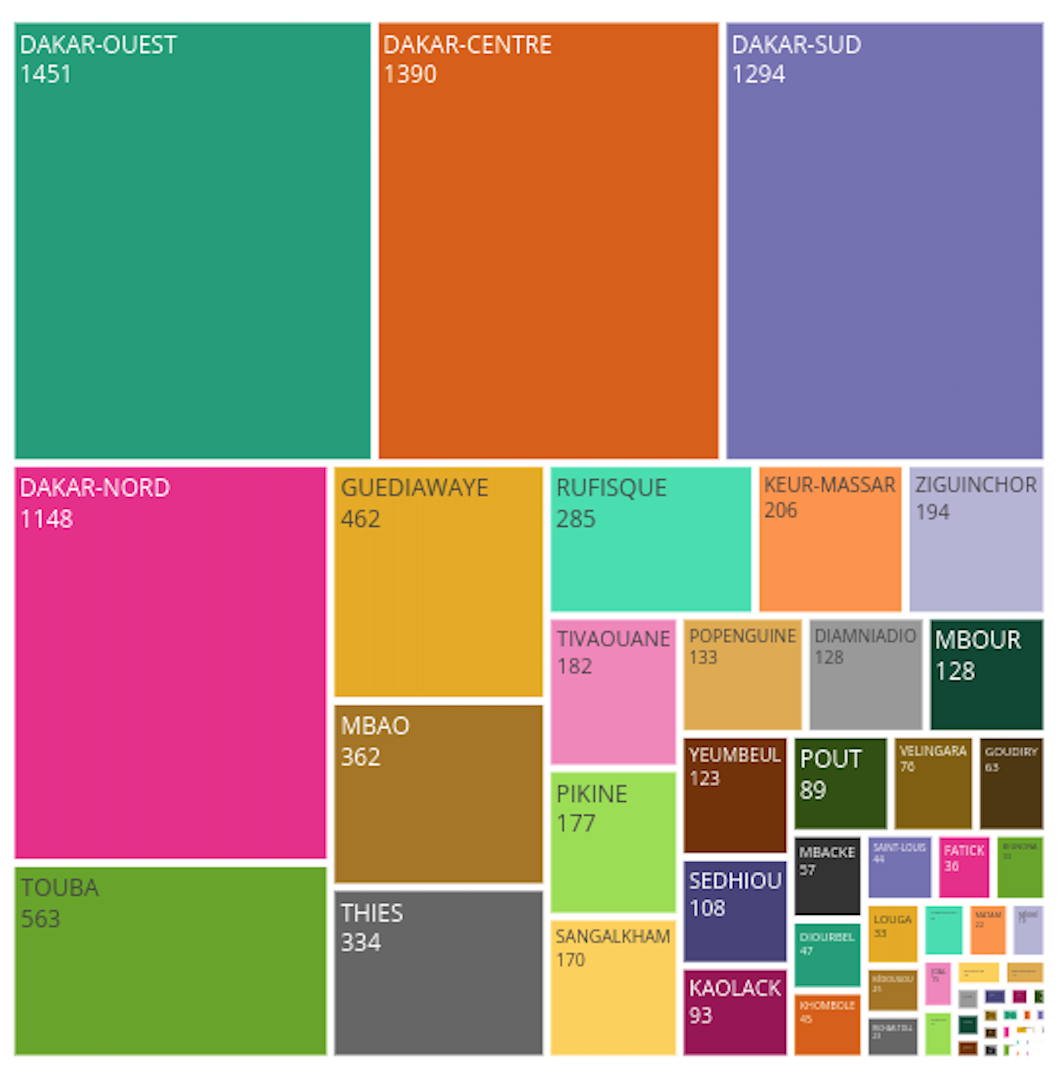}
	\par\vspace{-.25cm}
	\caption{Cumulative number of confirmed cases}\label{cumsumzones}
\end{figure}

If we take the distribution of contamination cases, we see that Dakar (with the high-density area) is the most contaminated. Figure \ref{dakar_viz} illustrates the visualization of high-density areas (for the North, South, West, and Center). Figure \ref{caseszones} shows this phenomenon with the first 5 (Figure \ref {first5_zones}) and the first 10 (Figure \ref{first10_zones}). Besides, in Figure \ref{cumsumzones}, we show the cumulative repartition of the number of confirmed cases per zone.

\begin{figure}[h!]
	\centering
	\includegraphics[width=0.7\linewidth]{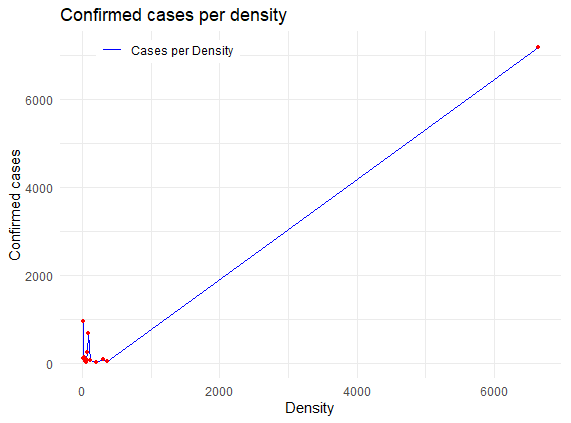}
	\par\vspace{-0.25cm}
	\caption{Confirmed cases per region density}\label{casedens}
\end{figure}

\begin{figure}[h!]
	\centering
	\includegraphics[width=0.65\linewidth]{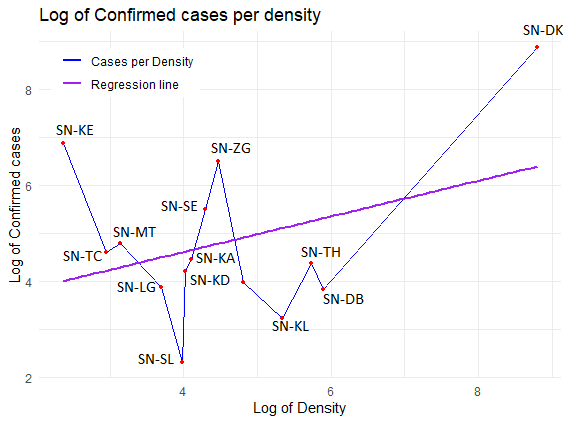}
		\par\vspace{-0.25cm}
	\caption{Logarithm of confirmed cases per logarithm of region density}\label{logconfreg}
\end{figure}

\begin{figure}[h!]
	\centering
	\includegraphics[width=0.8\linewidth]{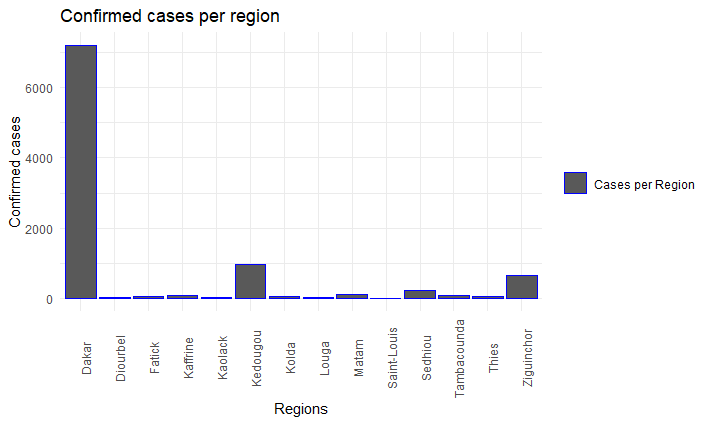}
		\par\vspace{-0.25cm}
	\caption{Histogram of confirmed cases per region}\label{casereg}
\end{figure}

\begin{figure}[h!]
	\centering
	\includegraphics[width=0.8\linewidth]{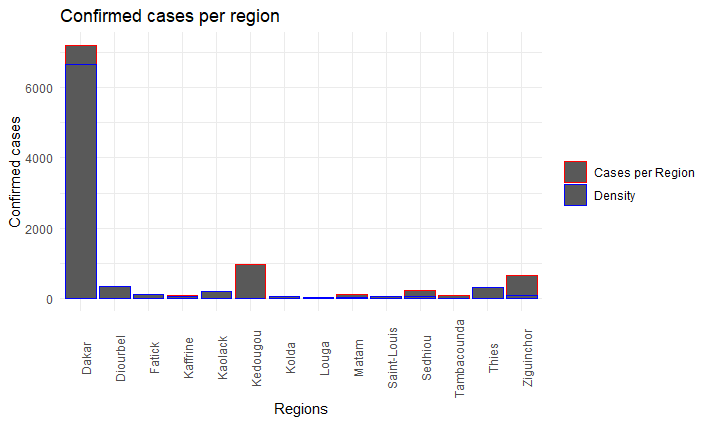}
		\par\vspace{-0.25cm}
	\caption{Histogram of confirmed cases and density per region}\label{casedensreg}
\end{figure}

\noindent High population densities can catalyze the spread of COVID-19 \cite{sarr}. With its 3,137,196 inhabitants, or almost a quarter of the population of Senegal (23.2\%), living on an area representing 0.3\% of the total area of the country, Dakar is the most populated region of Senegal, and its population density is also the highest with 5,846 people/km$^2$.\\
\noindent Parcelles Assainies, Guediewaye are zones with high density.  That involves community cases (see Figure \ref{dkrGPM}), where the urbanization rate is 44\%. By July 26, the active cases in Dakar are 5283  (West (1451), South  (1294),  North (1148), and Center (1390)). 
Keeping to more than one meter (1 m) distance between people coughing and sneezing, as recommended by the WHO, becomes more difficult with higher population densities like in Dakar. Therefore, avoiding situations with higher population densities will be a necessary requirement to limit the spread of COVID-19.\\
The Figure \ref{casedens} shows the confirmed cases by the density of population in each region, Figure \ref{logconfreg} shows the logarithm of confirmed cases per logarithm of density of population in each region, Figure \ref{casereg} shows the confirmed cases per region and Figure \ref{casedensreg} shows the confirmed cases and density per region. In  Figure \ref{logconfreg}, we use the ``CodesISO3166-2'' to characterize the regions: Dakar(SN-DK), Zinguinchor(SN-ZG), Diourbel(SN-DB), Saint-Louis(SN-SL), Tambacounda(SN-TC), Kaolack(SN-KL), Thies(SN-TH), Louga(SN-LG), Fatick(SN-FK), Kolda(SN-KD), Matam(SN-MT), Kaffrine(SN-KA), Kedougou(SN-KE), Sedhiou(SN-SE). 
We see that for some regions with high density, there are few cases,  while for others with low density, there are many cases. However, the Dakar region with the highest density has the most cases.

\subsection{Mortality, recovery rates and daily increase}
\begin{figure}[h!]
  \subfloat[recovery rate]{
	\begin{minipage}[1\width]{0.5\textwidth}
	   \centering
	   \includegraphics[width=1\textwidth]{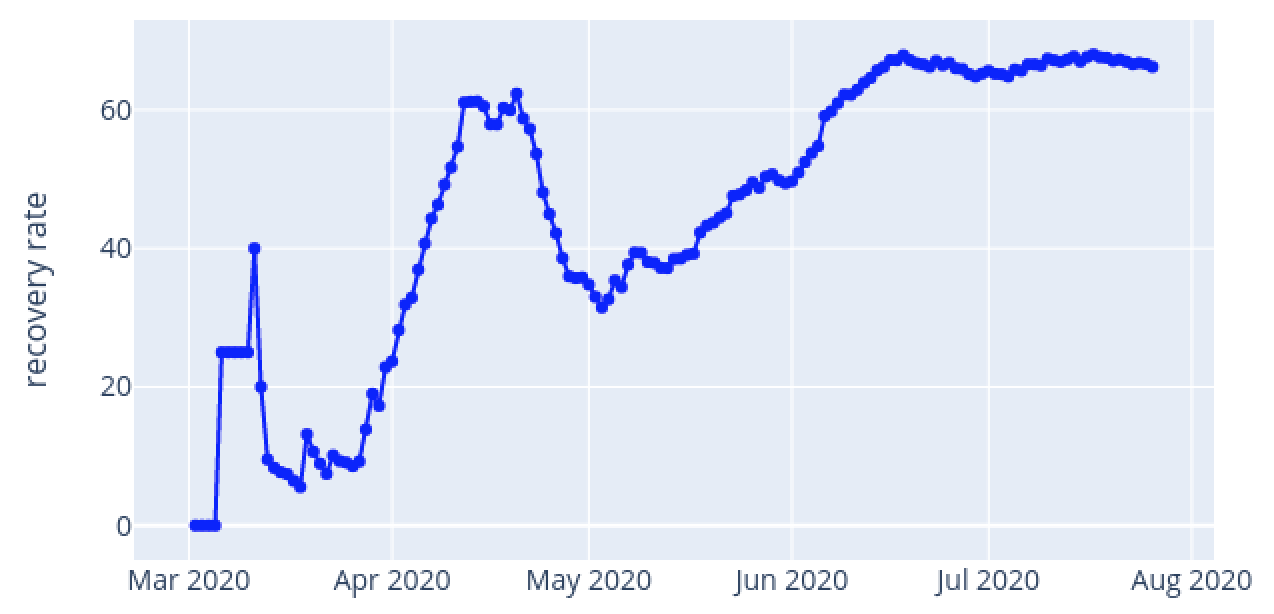}\label{recov_rate}
	\end{minipage}}
% \hfill 	
  \subfloat[mortality rate]{
	\begin{minipage}[1\width]{ 0.5\textwidth}
	   \centering
	   \includegraphics[width=1\textwidth]{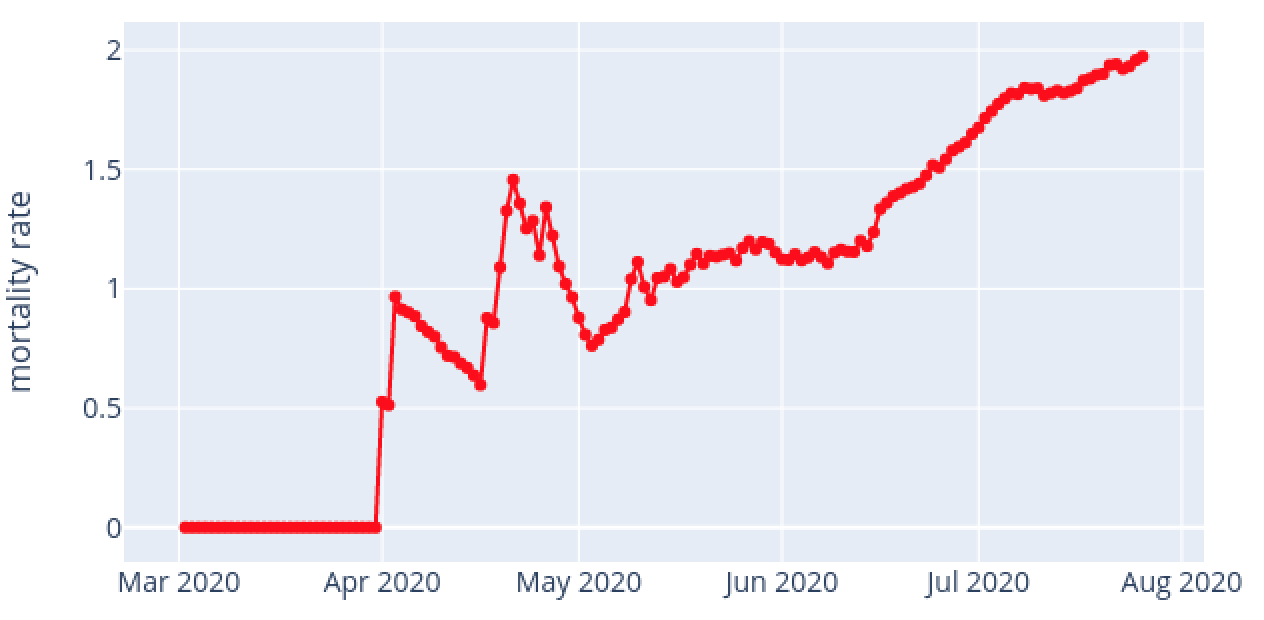}\label{mortal_rate}
	\end{minipage}}
\caption{Evolution of the recovery and mortality rates}\label{sn_recovmortrates}
\end{figure}

\begin{figure}[h!]
	\centering
	\includegraphics[width=0.9\linewidth]{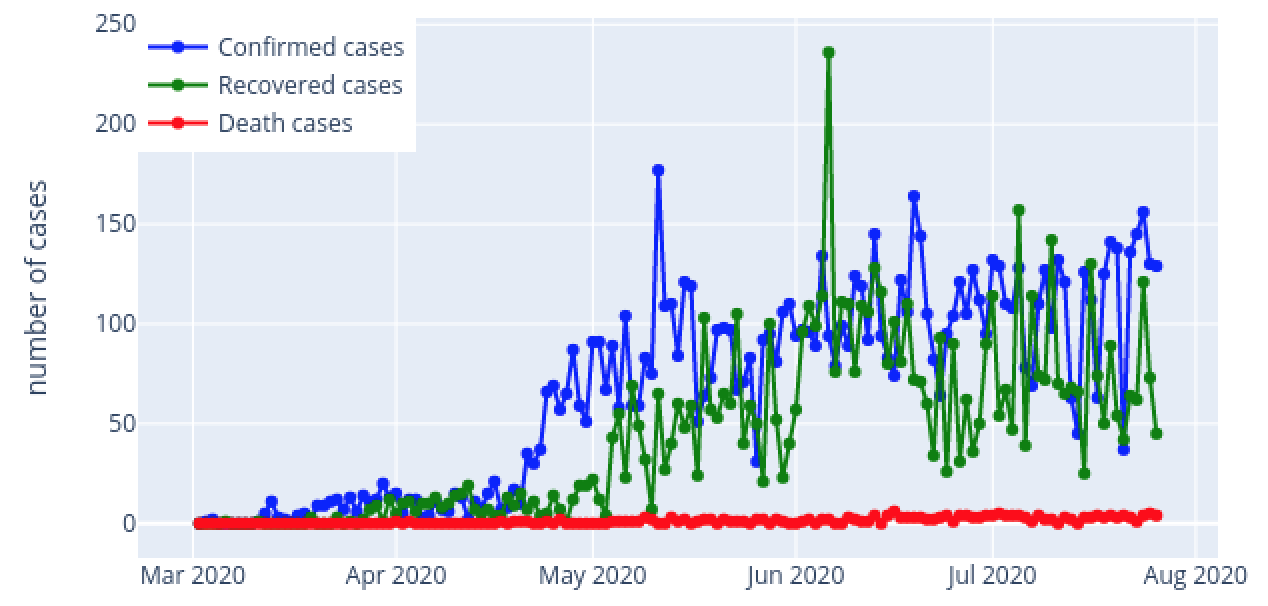}
		\par\vspace{-.25cm}
	\caption{Senegal: daily increase cases}\label{sn_dailyincrease}
\end{figure}

First, let's define the mortality and recovery rates. 
$$
\mbox{Mortality rate} = \frac{\mbox{number of death cases}}{\mbox{number of confirmed cases}} \times 100
$$
$$
\mbox{Recovery rate}= \frac{\mbox{number of recoverd cases}}{\mbox{number of confirmed cases}} \times 100
$$
The recovery and mortality rates are given by Figure \ref{sn_recovmortrates}, and their  average and median values in Table \ref{sn_avermedvalues}. \\
\begin{table}[h!] 
\begin{center}
\begin{tabular}{|l|c|} 
 \hline
 {\bf average/median} 	& {\bf values}  \\ \hline
average recovery rate &   45.760529  \\\hline
median recovery rate &   49.509649 \\\hline
\hline
average mortality rate &   1.006177  \\\hline
median mortality rate & 1.119125  \\\hline
\end{tabular}
\end{center}
\caption{Senegal: average and median values}\label{sn_avermedvalues}
\end{table}
\noindent We see that the recovery rate has started to pick up again, which is a good sign.
The daily increase of confirmed, recovered, and death of cases are given in Figure \ref{sn_dailyincrease}, and their average increase in Table \ref{sn_averincrease}.
\begin{table}[h!] 
\begin{center}
\begin{tabular}{|l|c|} 
 \hline
 {\bf average increase every day} 	& {\bf values}  \\ \hline
average increase of confirmed cases every day &  66.0 \\ \hline
average increase of recovered cases every day & 44.0 \\ \hline
average increase of deaths cases every day &  1.0\\ \hline
\end{tabular}
\end{center}
\caption{Senegal: average increase  cases }\label{sn_averincrease}
\end{table}

%-----%
\subsection{Growth factor}
\subsubsection{Growth factor for confirmed, recovered and deaths cases}
\begin{figure}[h!]
	\centering
	\includegraphics[width=1.\linewidth]{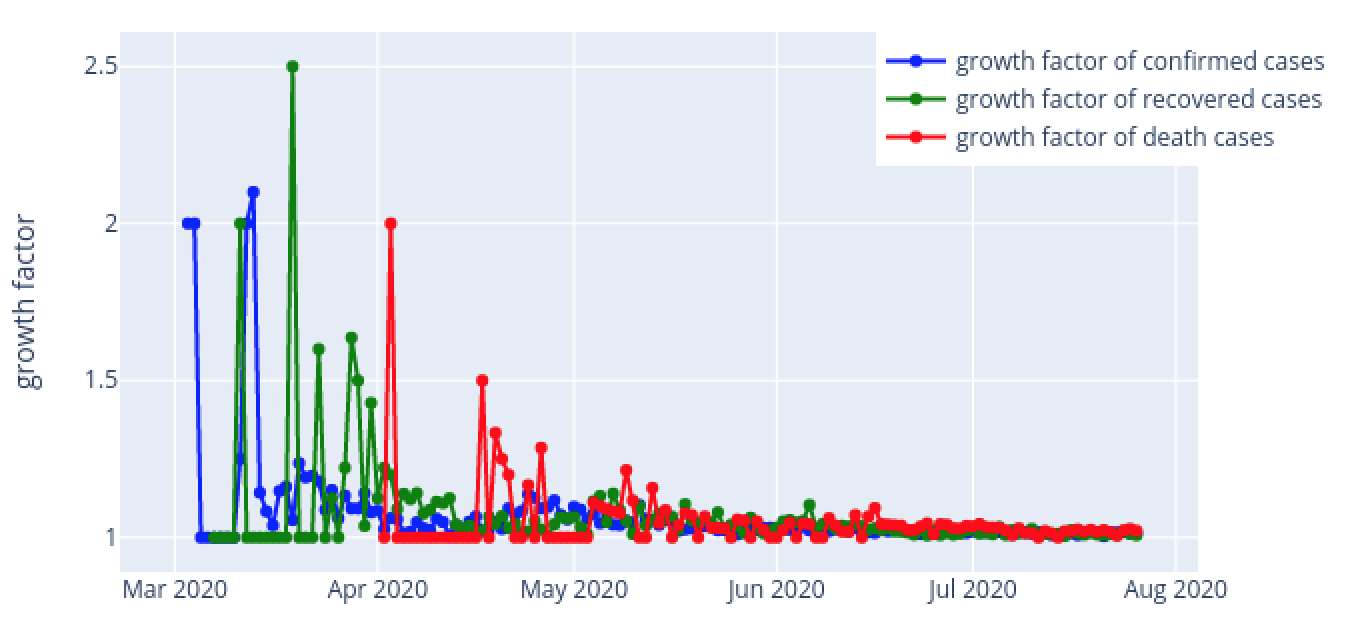}
		\par\vspace{-.25cm}
	\caption{Senegal: growth factor for  confirmed,  recovered and deaths cases}\label{sn_growfact}
\end{figure}

The growth factor is the factor by which a quantity multiplies itself over time. The formulas used are: \\
For confirmed cases:
$$
 \qquad \frac{\mbox{Every day's new Confirmed}}{\mbox{new Confirmed on the previous da}y}
$$
For recovered cases:
$$
 \qquad \frac{\mbox{Every day's new Recovered}}{\mbox{new Recovered on the previous da}y}
$$
For deaths cases:
$$
 \qquad \frac{\mbox{Every day's new Deaths}}{\mbox{new Deaths on the previous da}y}
$$
\begin{rem}
\begin{itemize}
\item A growth factor constant at 1 indicates there is no change in any kind of cases.
\item A growth factor above 1 indicates an increase corresponding cases.
\item A growth factor above 1 but trending downward is a positive sign, whereas a growth factor constantly above 1 is the sign of exponential growth.
\end{itemize}
\end{rem}
\noindent The growth factor for confirmed, recovered and deaths of cases is given by Figure \ref{sn_growfact}, and their average and median growth factors in Table \ref{sn_avermedincrease}.
\par\vspace{-0.35cm}
\begin{table}[h!] 
\begin{center}
\begin{tabular}{|l|c|} 
 \hline
 {\bf average/median growth factor} 	& {\bf values}  \\ \hline
average growth factor of number of confirmed cases & 1.073613  \\ \hline
median growth factor of number of confirmed cases &   1.027353 \\ \hline \hline
average growth factor of number of recovered cases &  inf \\ \hline
median growth factor of number of recovered cases & 1.029787  \\ \hline
\hline
average growth factor of number of death cases & inf \\ \hline
median growth factor of number of death cases & 1.023810 \\ \hline
\end{tabular}
\end{center}
\caption{Senegal: average  and median growth factors of cases }\label{sn_avermedincrease}
\end{table}

\subsubsection{Growth factor for active and closed cases}
\begin{figure}[h!]
	\centering
	\includegraphics[width=.95\linewidth]{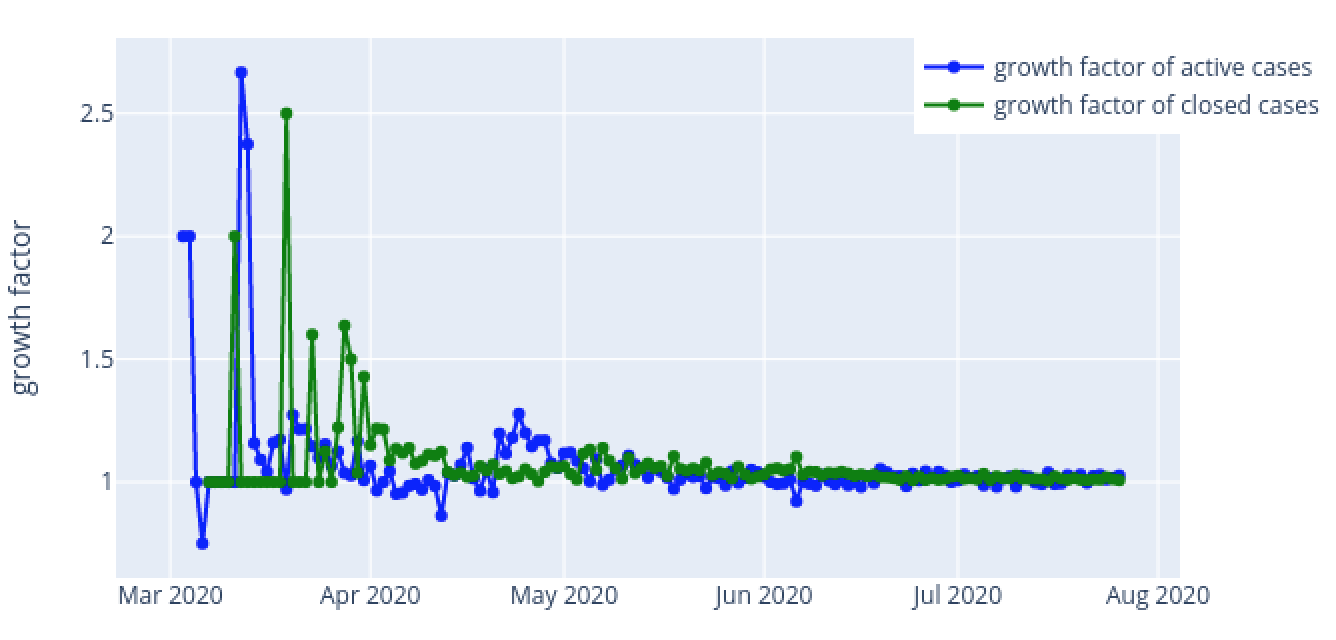}
		\par\vspace{-.25cm}
	\caption{Senegal: growth factor for  active and closed cases}\label{sn_growthfactorIC}
\end{figure}

The formulas used are: \\
For active cases:
$$
 \qquad \frac{\mbox{Every day's new Active}}{\mbox{new active on the previous da}y}
$$ 
For closed cases:
$$
 \qquad \frac{\mbox{Every day's new Closed}}{\mbox{new Closed on the previous da}y}
$$ 
The growth factor for active and closed cases is given by Figure \ref{sn_growthfactorIC}.

\subsection{Unreported cases}
\begin{figure}[h!]
	\centering
	\includegraphics[width=0.95\linewidth]{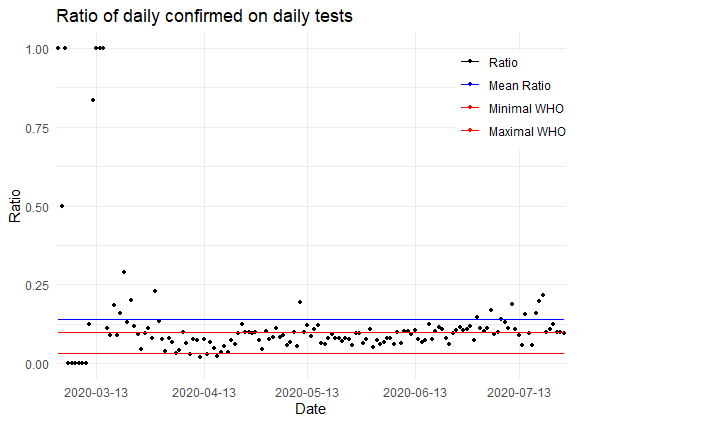}
	\par\vspace{-0.2cm}
	\caption{Senegal: ratio}\label{Ratiowho}
\end{figure}

\begin{figure}[h!]
	\centering
	\includegraphics[width=0.95\linewidth]{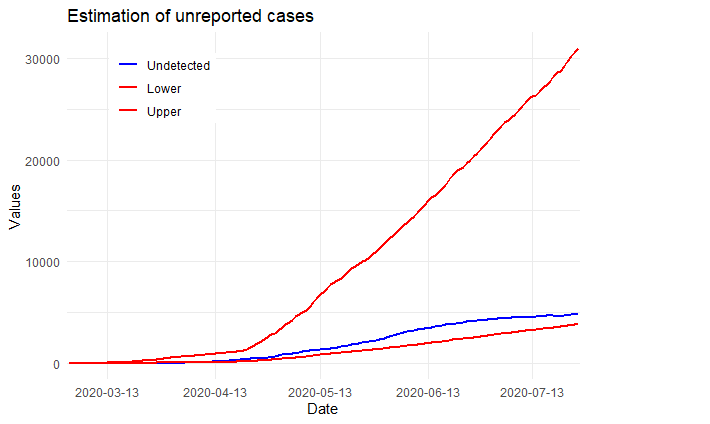}
	\par\vspace{-0.2cm}
	\caption{Senegal: estimation of unreported cases}\label{Unrep}
\end{figure}

We can see in Figure \ref{Ratiowho} that starting around March 13 to July 07, the ratios are around the mean 10\%. 
That means the number of tests influences the number of daily cases, as we see in Figures \ref{tests} of the number of tests and Figure \ref{sn_weeklyincrease} of weekly increase. \\ 
If we consider that the mean ratio shows the real situation of the country in the real number of infectious individuals, we can compute approximately the number of unreported cases in the country. We can use the formula $U=Ratio \ T - C$, where $T$ is the daily test or for simplifying the mean daily test, $C$ is the daily confirmed case.
We know by WHO that testing is the key to controlling the virus. The WHO considers that the more tests that are conducted, the easier it becomes to track the spread of the virus and reduce transmission. The WHO has suggested around $10$, $30$ tests per confirmed case as a general benchmark of adequate testing. \\
By considering that an excellent way to spread the virus is to perform several tests between $10 C$ and $30 C$, $C$ is the daily confirmed cases. That means the daily and mean ratio must be in the interval $[1/30,\ 1/10]$. In Figure \ref{Ratiowho}, we show the daily and mean ratio and the recommended lower and upper ratios. \\
It is clear that if the ratio is higher than the recommended upper ratio, there are undetected cases. Those cases have delay for contamination until they become confirmed (see Figure \ref{Unrep} for the estimation of unreported cases). 

%-------%
\section{Forecasting using machine learning models}\label{prediction}
In this section, first we perform 5 days ahead forecast with Linear regression \cite{fan}, Polynomial regression \cite{fan}, Support Vector Regression (SVR) \cite{Vapnik}, Prophet \cite{prophet} and Multilayer Perceptron (MLP) \cite{Alpaydin:2010,Freund-Shapire}. Second, we compare their Root Mean Square Error (RMSE). The RMSE measures how much error there is between two data sets. In other words, it compares a predicted value  (${{\widehat y}_i}$) and an observed or known value ($y$). The smaller the RMSE value is, the closer predicted and observed values are. 
Also, we make a forecasting (2 weeks and 40 days) with the two methods giving the two best RMSE (i.e., Prophet and MLP).  \\
Recall that the sum of squared errors is defined by: 
$$
\sum\limits_{i = 1}^n \left( {{y_i} - {{\widehat y}_i}} \right) ^2.
$$
The mean square error (MSE) is the arithmetic mean of the squares of the deviations between the model predictions and the observations, and the RMSE is the root of the MSE. The mean absolute error (MAE) is the arithmetic mean of the absolute values of the deviations. The mean absolute percentage error (MAPE) is the average of the deviations in absolute value from the observed values.\\
The MSE, RMSE, MAE and MAPE  formula are given, respectively, by:
$$
\frac{1}{n}\sum\limits_{i = 1}^n \left( {{y_i} - {{\widehat y}_i}} \right) ^2, \quad \sqrt{ \frac{1}{n}\sum\limits_{i = 1}^n \left( {{y_i} - {{\widehat y}_i}} \right) ^2}, \quad \frac{1}{n}\sum\limits_{i = 1}^n {\left| {{y_i} - {{\widehat y}_i}} \right|}, \quad \frac{1}{n}\sum\limits_{i = 1}^n 
{\left| \frac{  {{y_i} - {{\widehat y}_i}} }{y_i}\right|}
$$
In section \ref{diagnostic_comp} we compare with Prophet the MSE, RMSE, MAE and MAPE for a week forecasting, and in section \ref{comp_fore} the RMSE of the 5 proposed forecasting technics.

\subsection{Linear regression model}
\begin{figure}[h!]
	\centering
	\includegraphics[width=.9\linewidth]{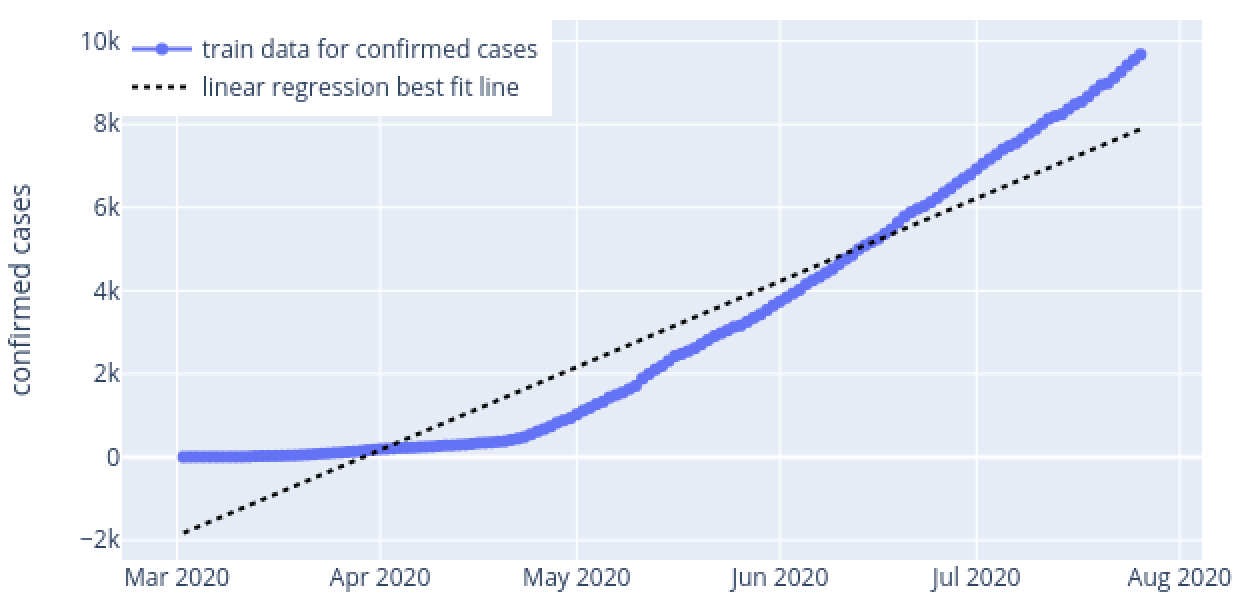}
	\par\vspace{-.25cm}
	\caption{Senegal: linear regression forecasting of confirmed cases}\label{sn_linearreg}
\end{figure}

Linear regression is the most straightforward and most widely used statistical technic for predictive modeling \cite{fan,bestfit}. 
Linear regression technics are used to create a linear model. The model describes the relationship between a dependent variable $y$  (also called the response) as a function of one or more independent variables $x_i$  (called the predictors). The general equation for a linear regression model is:
$$
y = w_0 + \sum_{i=1}^n w_ix_i +\epsilon_i
$$ 
where $w$  represents linear parameter estimates to be computed and $\epsilon$   the error terms.\\
\noindent There are several types of linear regression models: simple linear regression (model with only one predictor), multiple linear regression (model with multiple predictors), logistic regression, ordinal regression, multinomial regression, and discriminant analysis. Here, we are using simple linear regression.\\
\noindent The RMSE for linear regression is  1576.974603. The linear regression forecasting of confirmed cases is given in Figure \ref{sn_linearreg}. 
\noindent We see that the linear regression model is falling apart. As it is visible that the trend of confirmed cases in absolutely not linear. 
It cannot model the relationship between the target variable and the predictor variable.  In other words, they do not have a linear relationship. The polynomial regression might be of assistance.

\subsection{Polynomial Regression}
\begin{figure}[h!]
	\centering
	\includegraphics[width=.9\linewidth]{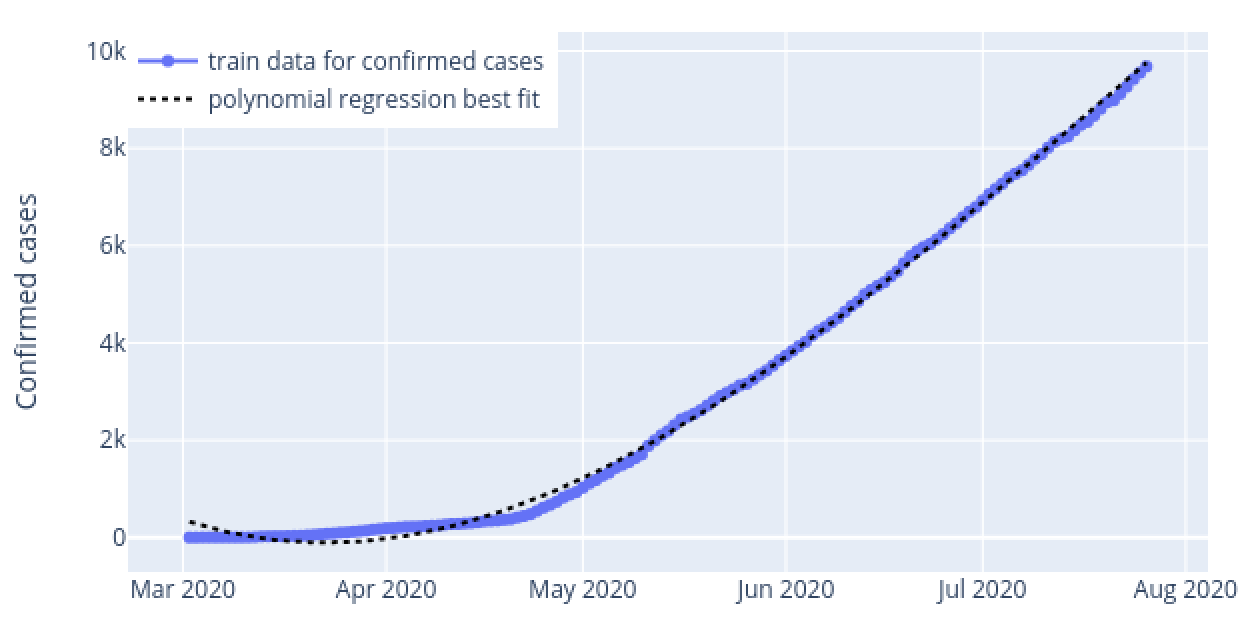}
	\par\vspace{-.25cm}
	\caption{Senegal: polynomial regression forecasting of confirmed cases}\label{sn_polyreg}
\end{figure}

Polynomial regression is another form of regression in which the maximum power of the independent variable is more than one \cite{fan,bestfit}. In this regression technic, the best fit line is not a straight line; instead, it is in the form of a curve. The advantages of using polynomial regression are: (i) polynomial fits a wide range of curvature, (ii) a broad range of function can be fitted under it, (iii) polynomial provides the best approximation of the relationship between the dependent and independent variable.\\
But, in polynomial regression, we have a polynomial equation of degree $n$ represented as:
$$
y = w_0 + w_1x + w_2x^2 + ... + w_nx^n
$$ 
where $x$ = score on the independent variable, $y$ = estimated dependent variable score, $w_0$ = constant and $w_1, w_2,..., w_n$ = are the weights in the equation of the polynomial regression and $n$ is the degree of the polynomial.\\
We can choose the degree of polynomial based on the relationship between target and predictor. The 1-degree polynomial is a simple linear regression; therefore, the value of degree must be greater than 1. Here, we have taken a 3-degree polynomial.\\
The RMSE for polynomial regression is 132.4346590595342. The polynomial regression forecasting of confirmed cases is given in Figure \ref{sn_polyreg}.
We can observe that polynomial regression is better at fitting the data than linear regression. Also, due to better-fitting, the RMSE of polynomial regression is way lower than that of linear regression.
\subsection{Support Vector Machine - Regression (SVR)}
\begin{figure}[h!]
	\centering
	\includegraphics[width=.9\linewidth]{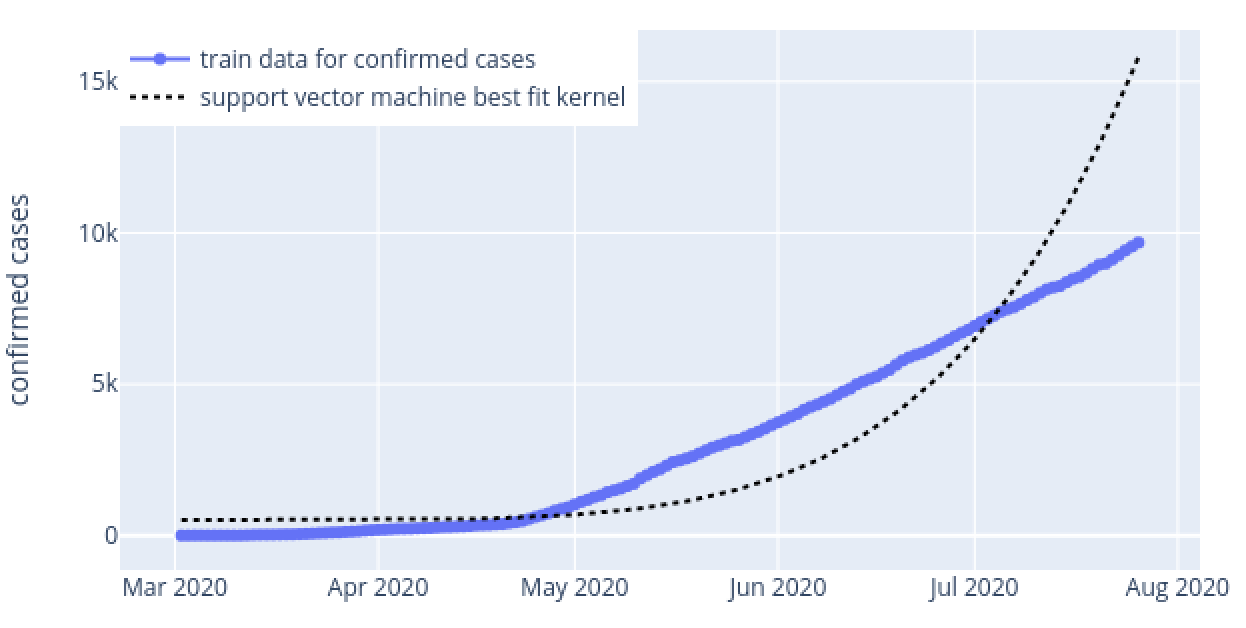}
	\par\vspace{-.25cm}
	\caption{Senegal: SVM forecasting of confirmed cases}\label{sn_svm}
\end{figure}

Support Vector Machine (SVM) is a supervised machine learning algorithm that can be used for both classification or regression challenges \cite{Vapnik}. 
As a regression method, it maintains all the main features that characterize the algorithm (maximal margin). The Support Vector Regression (SVR) uses the same principles as the SVM for classification, with only a few minor differences. SVM regression is considered a nonparametric technic because it relies on kernel functions. 
The goal is to find a function $f(x)$ that deviates from $y_n$ by value no greater than $\epsilon$ for each training point $x$, and at the same time is as flat as possible.\\
Suppose we have a set of training data where $x_i$ is a multivariate set of $n$ observations with observed response values $y_i$. To find the linear function $f(x)=xw+b$, and ensure that it is as flat as possible, find $f(x$) with the minimal norm value $||w||^2$. This is formulated as a convex optimization problem to minimize:
\begin{align*}
\min  \  & \ \frac{1}{2} ||w||^2 \\
s.t.\ & |y_i - (x_iw + b)|  \leq \epsilon \quad \forall i
\end{align*}
The function used to predict new values depends only on the support vectors: 
$$
y = \sum_{i=1}^n (\alpha_i-\alpha_i^*).(x_i,x) +b
$$
$\alpha_i$ and  $\alpha^*$  are nonnegative multipliers for each observation $x_i$.\\
The RMSE for SVR is: 4942.716486, and the SVR forecasting of confirmed cases is given by Figure \ref{sn_svm}.
In addition, we can clearly observe that polynomial regression is better at fitting the data than linear and SVM  regressions.

\subsection{Prophet model}
Prophet \cite{prophet} is a procedure for forecasting time series data based on an additive model where non-linear trends are fit with yearly, weekly, and daily seasonality, plus holiday effects. It works best with time series that have strong seasonal effects and several seasons of historical data. Prophet is robust to missing data and shifts in the trend, and typically handles outliers well. 
For the average method, the forecasts of all future values are equal to the average (or “mean”) of the historical data. If we let the historical data be denoted by $y_1,...,y_T$, then we can write the forecasts as
$$
\hat{y}_{T+h|T}=\bar{y}=(y_1+y_2+...+y_T)/T
$$
The notation $\hat{y}_{T+h|T}$ is a short-hand for the estimate of $y_{T+h}$  based on the data $y_1,...,y_T$.\\
A forecasting interval gives an interval within which we expect $y_t$  to lie with a specified probability. For example, assuming that the forecast errors follow a normal distribution, a 95\% forecasting interval for the  $h$-step forecast is 
$$
\hat{y}_{T+h|T}\pm1.96\hat{\sigma_h}
$$
where  ${\sigma_h}$ is an estimate of the standard deviation of the $h$-step forecast distribution.  
\subsubsection{Diagnostics}\label{diagnostic_comp}
Here, we make some diagnostics by using the cross validation (see Table \ref{sn_crossvalid}) and the performance metrics (see Table \ref{sn_perfmetrics}) using MSE, RMSE, MAE and MAPE. The Figure \ref{sn_cvm} illustrates these cross validation metrics, making 7 forecasts with cutoffs between 2020-07-15, 00:00:00 and 2020-07-21, 00:00:00 (initial='135days', period='1 days', horizon = '5 days'). 
\par\vspace{-0.5cm}
\begin{figure}[h!]
  \subfloat[mape]{
	\begin{minipage}[1\width]{0.47\textwidth}
	   \centering
	   \includegraphics[width=1.\textwidth]{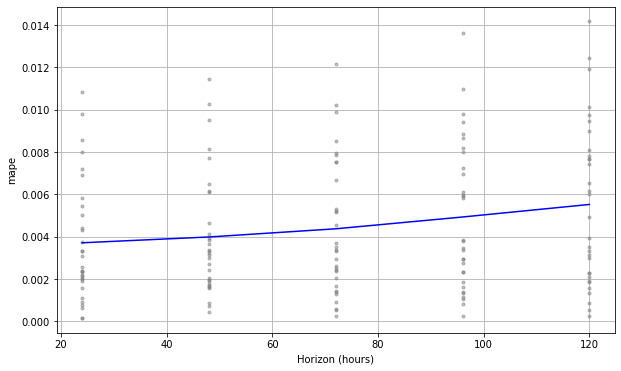}\label{sn_mape}
	\end{minipage}}
% \hfill 	
  \subfloat[mae]{
	\begin{minipage}[1\width]{ 0.47\textwidth}
	   \centering
	   \includegraphics[width=1.\textwidth]{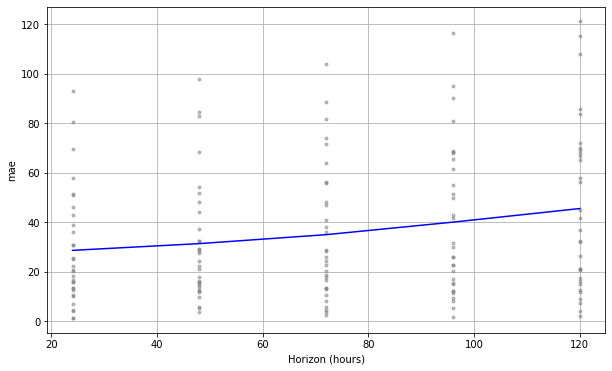}\label{sn_mae}
	\end{minipage}}
\newline
  \subfloat[rmse]{
	\begin{minipage}[1\width]{ 0.47\textwidth}
	   \centering
	   \includegraphics[width=1.\textwidth]{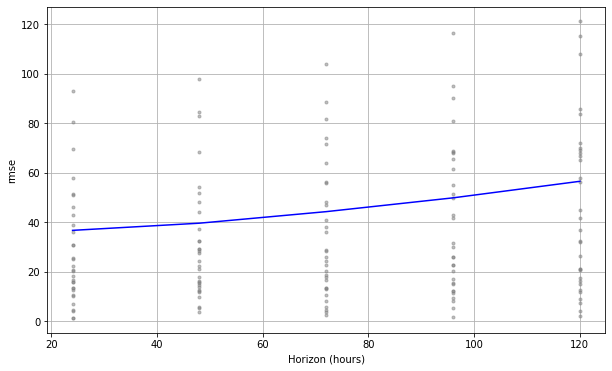}\label{sn_rmse}
	\end{minipage}}
%
% \hfill 	
  \subfloat[mse]{
	\begin{minipage}[1\width]{ 0.47\textwidth}
	   \centering
	   \includegraphics[width=1.\textwidth]{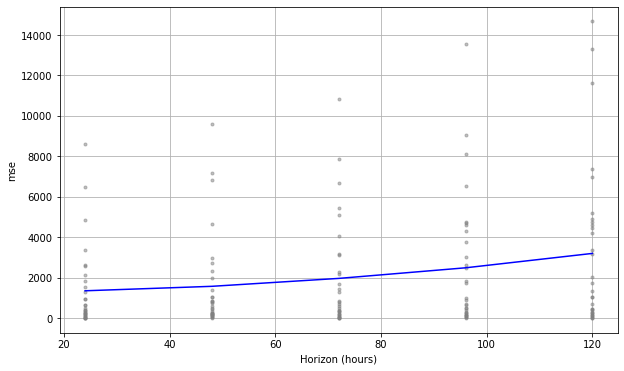}\label{sn_mse}
	\end{minipage}}
	\caption{Senegal: cross validation metrics}\label{sn_cvm}
\end{figure}
\begin{table}[h!] 
\begin{center}
\begin{tabular}{|c|c|c|c|c|c|} 
 \hline
          {\bf ds}   &         ${\bf \hat{y}}$ &    ${\bf\hat{y}_{lower}}$ &    ${\bf\hat{y}_{upper}}$  & ${\bf y} $	& {\bf cutoff} \\  \hline
2020-07-16	& 8527.102656&	8490.194455&	8559.523867&	8481&	2020-07-15 \\  \hline
	2020-07-17&	8642.037561&	8604.002596&	8681.562780&	8544&	2020-07-15\\  \hline
	2020-07-18&	8757.731562&	8713.796517&	8801.281978&	8669&	2020-07-15\\  \hline
	2020-07-19	&8861.406842&	8806.387035&	8922.864695&	8810&	2020-07-15\\  \hline
	2020-07-20	& 8965.288487&	8892.653796&	9039.413469&	8948&	2020-07-15\\  \hline
\end{tabular}
\end{center}
\caption{Senegal: cross validation}\label{sn_crossvalid}
\end{table}
\begin{table}[h!] 
\begin{center}
\begin{tabular}{|c|c|c|c|c|} 
 \hline
 	{\bf horizon} & {\bf mse } & {\bf rmse} 	&{\bf mae} &	{\bf mape}  \\  \hline
		1 day  &	1346.372683	&36.692951	&28.612794 	& 0.003718\\  \hline
	2 days&	1567.467031& 	39.591249&	31.337141&	0.003992\\  \hline
	3 days& 	1958.774371&	44.258043&	34.955291& 	0.004379\\  \hline
	4 days &	2483.073134& 	49.830444& 	40.018887& 	0.004936\\  \hline
	5 days &	3194.413073&	56.519139& 	45.507174& 	0.005528\\  \hline	
\end{tabular}
\end{center}
\caption{Senegal: performance metrics with Prophet}\label{sn_perfmetrics}
\end{table}
\noindent From Table \ref{sn_crossvalid}, by comparing the values obtained on July 20, 2020 (column $y$=8948) with the predicted one (column $ \hat{y}$=8965.288487), we see that the error is 0.19\%. The predicted value is always within the confidence interval. So, Prophet seems to give us good value.

\subsubsection{Trend changepoints and forecasting}
\begin{figure}[h!]
	\centering
	\includegraphics[width=.9\linewidth]{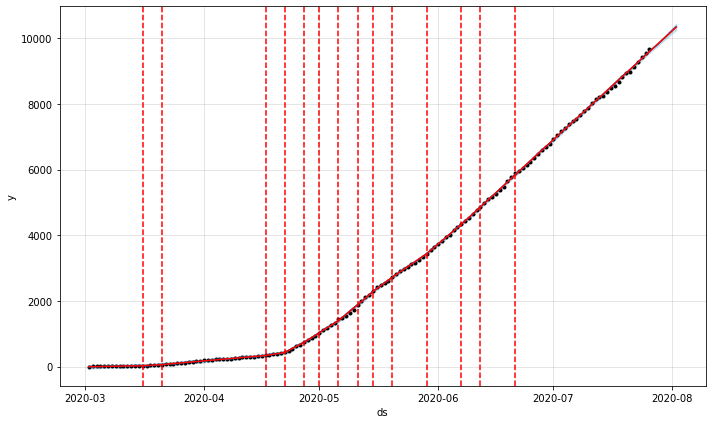}
	\caption{Senegal: changepoints of confirmed cases}\label{sn_changepoints}
\end{figure}

\begin{figure}[h!]
	\centering
	\includegraphics[width=.85\linewidth]{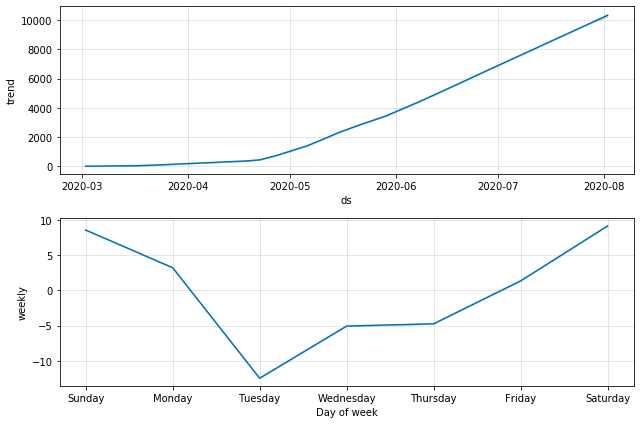}
	\caption{Senegal: Trends and weekly indrease of confirmed cases}\label{sn_procomponents}
\end{figure}

\begin{figure}[h!]
  \subfloat[2 weeks forecasting]{
	\begin{minipage}[1\width]{0.47\textwidth}
	   \centering
	   \includegraphics[width=1.1\textwidth]{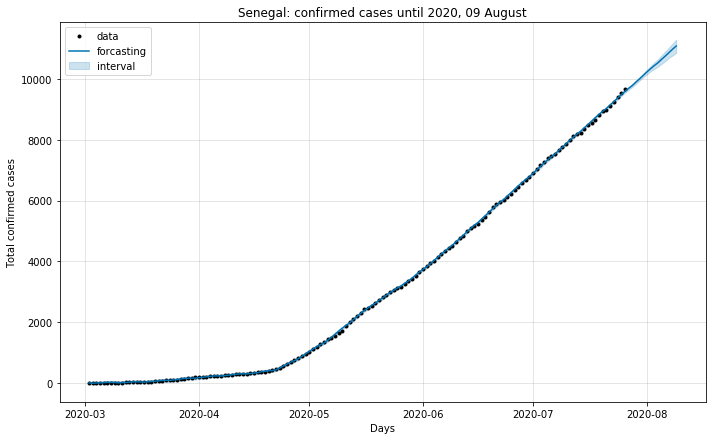}\label{prophet_2weeks_fore}
	\end{minipage}}
% \hfill 	
  \subfloat[40 days forecasting]{
	\begin{minipage}[1\width]{ 0.47\textwidth}
	   \centering
	   \includegraphics[width=1.1\textwidth]{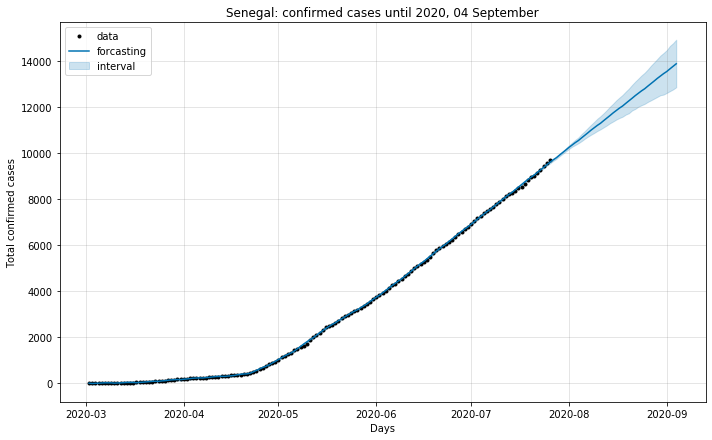}\label{prophet_40days_fore}
	\end{minipage}}
	\caption{Senegal: Prophet for forecasting of confirmed cases}\label{sn_cvm}
\end{figure}

The RMSE for Prophet model is  24.332935. The Prophet forecasting of confirmed cases, with trend changepoints, is given by Figure \ref{sn_changepoints}, and the trends and weekly increase are given by Figure \ref{sn_procomponents}. With Prophet, at $\sim$ August 09, 2020 we may obtain $>$ 11090 confirmed cases and  $>$ 13870 confirmed cases at $\sim$ September 04, 2020 (see Tables \ref{sn_1w_confcases} and \ref{sn_40d_confcases}). The forecasts of confirmed cases are illustrated in Figures \ref{prophet_2weeks_fore} and \ref{prophet_40days_fore}.\\
\begin{table}[h!]
\begin{center}
\begin{tabular}{|c|c|c|c|} 
 \hline
     {\bf ds}   &         ${\bf \hat{y}}$ &    ${\bf\hat{y}_{lower}}$ &    ${\bf\hat{y}_{upper}}$ \\  
           \hline
2020-08-05&	10655.486693& 	10510.221518&	10790.299094\\
           \hline
	2020-08-06&	10763.013788& 	10600.112043&	10914.058841\\
           \hline
	2020-08-07&	10876.303087& 	10707.775556&	11039.098323\\
           \hline
	2020-08-08&	10991.262629&	10782.447924&	11179.373374\\
           \hline
	2020-08-09&	11097.823573&	10875.493754&	11306.275975\\
           \hline
\end{tabular}
\end{center}
\caption{Prophet: predicted cumulative confirmed cases $\sim$August 09, 2020.}\label{sn_1w_confcases}
\end{table} 
\begin{table}[h!]
\begin{center}
\begin{tabular}{|c|c|c|c|} 
 \hline
 {\bf ds}   &         ${\bf \hat{y}}$ &    ${\bf\hat{y}_{lower}}$ &    ${\bf\hat{y}_{upper}}$ \\  
           \hline
	2020-08-31 &	13449.712068&	12538.804551&	14346.331331\\ \hline
	2020-09-01	& 13541.159323&	12611.551136&	14470.374884\\ \hline
	2020-09-02&	13655.655402&	12686.931428&	14650.954431\\ \hline
	2020-09-03&	13763.182496&	12762.246012&	14770.441020\\ \hline
	2020-09-04&	13876.471796&	12857.301393&	14925.442430\\ \hline
\end{tabular}
\end{center}
\caption{Prophet: predicted cumulative confirmed cases $\sim$September 04, 2020.}\label{sn_40d_confcases}
\end{table} 

\subsection{Artificial neural network}\label{ann}
\begin{figure}[h!]
	\centering
	\includegraphics[width=0.7\linewidth]{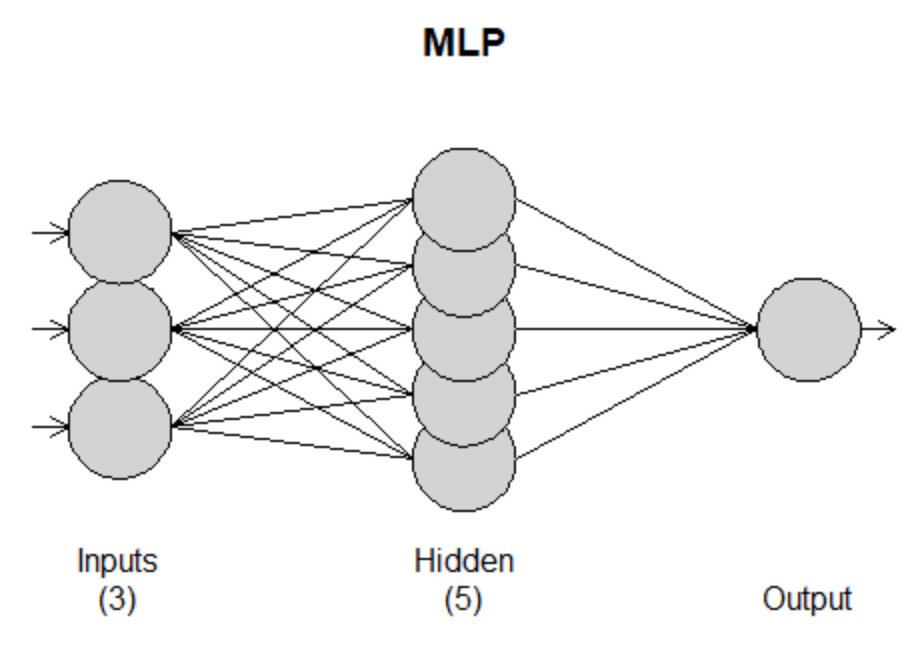}
	\par\vspace{-.25cm}
	\caption{Senegal: MLP fit with 5 hidden nodes and 20 repetitions.}\label{MLP}
\end{figure}

\begin{figure}[h!]
  \subfloat[2 weeks forecasting]{
	\begin{minipage}[1\width]{0.47\textwidth}
	   \centering
	   \includegraphics[width=1.1\textwidth]{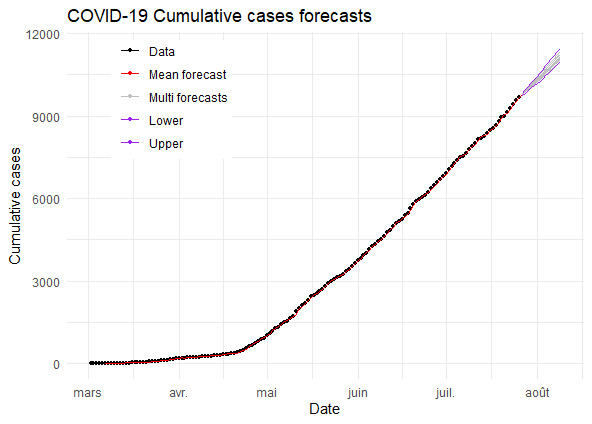}\label{mlp_2weeks_fore}
	\end{minipage}}
% \hfill 	
  \subfloat[40 days forecasting]{
	\begin{minipage}[1\width]{ 0.47\textwidth}
	   \centering
	   \includegraphics[width=1.1\textwidth]{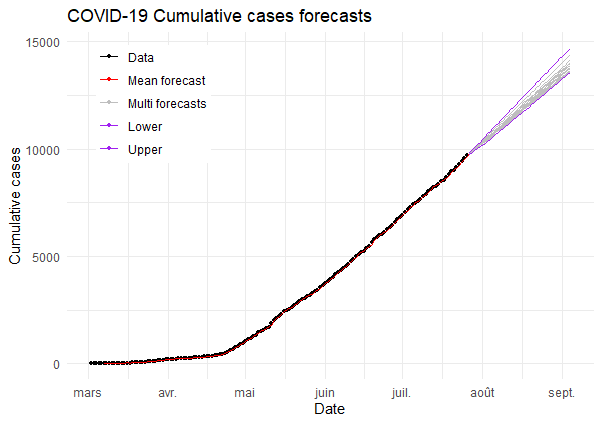}\label{mlp_40days_fore}
	\end{minipage}}
\caption{Senegal: MLP for forecasting of confirmed cases}\label{sn_MLP}
\end{figure}

Artificial neural networks are part of artificial intelligence. Biological neural networks are part of the animal brain. One of the main functions of the brain is to process information, and the primary information processing element is the neuron. This specialized brain cell combines (usually) several inputs to generate a single output. Depending on the animal, an entire brain can contain anywhere from a handful of neurons to more than a hundred billion, wired together. The output of one cell feeding the input of another, to create a neural network capable of remarkable feats of calculation and decision making (see \cite{Newman:2010}). 
If we could qualify the brain as a computer, then we would say that it is the best of computers. For this reason, the engineer seeks to improve mechanical computers to be closer to the biological computer, i.e., the brain.
The more neural connections there are, the more the network can solve complex problems. Pattern recognition is a task that neural networks can easily accomplish. For this task, introducing as input a pattern to a neural network, yields as output a pattern back (see \cite{Heaton:2015}). \\
In general, neural network problems involve a dataset used to predict values for later datasets. For that, the neural network needs to be trained. Then, neural networks can predict the outcome of entirely new datasets based on training from old data sets. 
Most neural network structures use some type of neuron, node, or unit. An algorithm called a neural network would generally be made up of individual interconnected neurons. \\
The artificial neuron receives input from one or more sources, which may be other neurons or data entered into the network from a computer program (see Figure \ref{MLP}). This entry is usually a floating-point or binary. Often the binary input is coded floating point representing true or false like $ 1 $ or $ 0 $. Sometimes the program also describes the binary input as using a bipolar system with true as $ 1 $ and false as $ -1 $.
An artificial neuron multiplies each of these inputs by a weight. It then adds these multiplications and transmits this sum to an activation function given by:
\begin{equation}
f(x_{i},w_{i})=\phi(\sum_{i=1}^{n}w_{i}\cdot x_{i}),
\label{eq:af}
\end{equation}
with the variables $ x $ and $ w $ represent the input and the weights of the neuron, $n$ is the number of input and weight.\\
There is much class of Artificial neural networks, and each of them may be subdivided into class again. A feedforward neural network is a class ANN where connections between the nodes do not form a cycle. A multilayer perceptron (MLP) is a class of feedforward artificial neural network. An MLP may refer to networks composed of multiple layers of perceptrons, and a perceptron is an algorithm for supervised learning of binary classifiers (see \cite{Alpaydin:2010, GoodfellowBengioCourville:2016,Heaton:2015,Newman:2010}).
We use the nnfor R package, which allows time series forecasting with Multilayer Perceptrons (MLP) and Extreme Learning Machines (ELM). It relies on the neuralnet package for R \cite{CRAN}, which provides all the machinery to train MLPs (see \cite{Kourentzes:2019}).  
With MLP, at $\sim$ August 09, 2020 we may obtain $>$ 11110 confirmed cases and  $>$ 13790 confirmed cases at $\sim$ September 04, 2020 (see Tables \ref{data09-08-2020} and \ref{data04-09-2020}). The forecasts of confirmed cases are illustrated in Figures \ref{mlp_2weeks_fore} and \ref{mlp_40days_fore}.\\
	\begin{table}[h!]
		\begin{center}
			\begin{tabular}{|c|c|c|c|} 
				\hline
 {\bf Date}   &         ${\bf P}$ &    ${\bf P_{lower}}$ &    ${\bf P_{upper}}$ \\  
				\hline
				2020-08-05&	10716.477&	10558.952&	10957.686\\\hline
				2020-08-06&	10814.215&	10661.085&	11082.901\\\hline
				2020-08-07&	10911.786&	10766.739&	11208.026\\\hline
				2020-08-08&	11009.281&	10873.242&	11333.900\\\hline
				2020-08-09&	11110.988&	10965.396&	11459.136\\\hline
			\end{tabular}
			\end{center}
	\caption{MLP: forecasting 2 weeks (until 2020, 09 August).}\label{data09-08-2020} 
	\end{table}
\begin{table}[h!]
	\begin{center}
		\begin{tabular}{|c|c|c|c|} 
			\hline
 {\bf Date}   &         ${\bf P}$ &    ${\bf P_{lower}}$ &    ${\bf P_{upper}}$ \\  
			\hline
			2020-08-31&	13375.467&	13169.586&	14198.702\\\hline
			2020-09-01&	13481.011&	13270.426&	14322.128\\\hline
			2020-09-02&	13586.572&	13371.284&	14445.321\\\hline
			2020-09-03&	13692.030&	13472.156&	14568.216\\\hline
			2020-09-04&	13797.109&	13573.002&	14690.702\\\hline
		\end{tabular}
	\end{center}
		\caption{MLP: forecasting 40 days (until 2020, 04 September).}\label{data04-09-2020} 
\end{table}

%---------------------%
\subsection{Comments and summary of forecasts using different models}\label{comp_fore}
\noindent  The table \ref{sn_comp_rmse} gives all rmse values obtained from the 5 methods. 
\noindent The confirmed cases forecasting with linear regression, polynomial regression, SVR, Prophet and Multilayer perceptron are given by Table \ref{sn_compare_all}.\\
We clearly observe that polynomial regression is better at fitting the data than linear regression and SVR. Also, based on the cross-validation and RMSE, MLP (the best fitting) and Prophet are better at fitting the data than linear, polynomial, and SVR.
 \begin{table}[h!] 
\begin{center}
\begin{tabular}{|l|c|} 
 \hline
  {\bf  Model name} 	& {\bf Root Mean Squared Error (RMSE)} \\ \hline
	Linear Regression & 	1576.974603 \\ \hline
	Polynomial Regression &	132.434659\\ \hline
 	Support Vector Machine Regressor &	4942.716486 \\ \hline
      	Prophet Model   &	24.332935\\ \hline
      	Mutilayer Perceptron & 19.71207\\ \hline 
      	\end{tabular}
\end{center}
\caption{Senegal: root mean square error for all models}\label{sn_comp_rmse}
\end{table}
\par\vspace{-1.cm}
 \begin{table}[h!] 
\begin{center}
\begin{tabular}{|c|c|c|c|c|c|} 
 \hline
{\bf Date} &	{\bf Linear} &  	{\bf Polynomial}   &	{\bf SVR}   & {\bf  MLP}   & {\bf	Prophet} \\ \hline
 	2020-07-27&	7951.692390&	9872.025007&	16345.262343& 9798.7073  &  	9699.794909	\\ \hline	
2020-07-28	& 8018.221190& 	9987.827481&	16891.321361&  9905.9566 &  	9791.304248 	\\ \hline	
2020-07-29	&  8084.749989&	10103.531274&	17452.339861&   10015.025 & 	 9905.861512  	\\ \hline	
2020-07-30&	8151.278788&	10219.120562&	18028.623125&10121.759    & 10013.346832  	\\ \hline		
2020-07-31&	8217.807587&	10334.579520&	18620.480560&  10222.750 & 10126.594134 	\\ \hline	
\end{tabular}
\end{center}
\caption{Senegal: comparison values for confirmed cases forecasting}\label{sn_compare_all}
\end{table}
\par\vspace{-1.cm}

%---------------------%
\section{Conclusion and perspectives}\label{ccl}
 Many questions deserve to be raised \cite {grid}: pharmacopeia to the rescue of a broken down modern medicine, biological diagnosis, some prevention strategies, the resilience of the Senegalese economy, and the legal consequences of COVID-19.
Forecasts show that the number of contamination in Senegal continues to climb despite the measures taken by the government. The weeks 16, 18 and 21 were very particular because there are the top in deaths and confirmed cases.\\
Severe measures still need to be put in place to reach the peak. Measures could help for the reopening of schools and universities, and hope to be among the countries authorized by the EU for their airspace.\\

\subsection*{Acknowledgement}
The authors thanks the Non Linear Analysis, Geometry and Applications (NLAGA) project for supporting this work (\url{http://nlaga-simons.ucad.sn}). 

%

% ------------------------------------------------------------------------%

\begin{thebibliography}{10}
%
\bibitem{ndiaye1} 
B.M. Ndiaye, L. Tendeng, D. Seck, Analysis of the COVID-19 pandemic by SIR model and machine learning technics for forecasting, arXiv:2004.01574v1 [q-bio.PE], 3 Apr 2020, \url{https://arxiv.org/pdf/2004.01574.pdf}.
%
\bibitem{ndiaye2}  
B.M. Ndiaye, L. Tendeng, D. Seck, Comparative prediction of confirmed cases with COVID-19 pandemic by machine learning, deterministic and stochastic SIR models, arXiv:2004.13489 [q-bio.PE], 24 Apr 2020, \url{https://arxiv.org/pdf/2004.13489.pdf}. 
%
\bibitem{ndiaye3} 
M.A.M.T. Balde, C. Balde, B.M. Ndiaye, Impact studies of nationwide measures COVID-19 anti-pandemic: compartmental model and machine learning,
arXiv:2005.08395 [q-bio.PE], 17 May 2020. \url{https://arxiv.org/pdf/2005.08395.pdf}.
%
\bibitem{balde}
M.A.M.T. Balde, Fitting SIR model to COVID-19 pandemic data and comparative forecasting with machine learning, medRxiv preprint doi: \url{https://doi.org/10.1101/2020.04.26.20081042. (2010)}.
%
\bibitem{sarr}
V.M. Ndiaye, S.O. Sarr, B.M. Ndiaye, Impact of contamination factors on the COVID-19 evolution in Senegal, arXiv:2006.16326 [q-bio.PE], 29 Jun 2020, \url{https://arxiv.org/pdf/2006.16326.pdf}.
%
\bibitem{grid} 
S.O. SARR, A. Ndiaye, A. Badiane, M. Diouf, S.B. Lo, A.S. Badji, P.I. Ndiaye, B.M. Ndiaye, A. Kane, M.P. Sarr, Groupe de Recherche Interdisciplinaire pour le Developpement (GRID), Rapport interimaire n2 du 10 Juillet 2020. 
%
\bibitem{Alpaydin:2010}
E. Alpaydin, Introduction to Machine Learning 2nd ed, 584. Adaptive Computation and Machine Learning, (2010).
%
\bibitem{Freund-Shapire}
Y. Freund and R. E. Schapire. Large margin classification using perceptron algorithm. Machine Learning, 37(3):277-296, 1999.
%
\bibitem{GoodfellowBengioCourville:2016}
I. Goodfellow, Y. Bengio and A. Courville, Deep Learning, MIT Press. \url{http://www.deeplearningbook.org} (2016).
%
\bibitem{Heaton:2015}
J. Heaton, AIFH Volume 3: Deep Learning and Neural Networks, Heaton Research, Inc, 268. Tracy Heaton (2015).
%
\bibitem{fan}
Fan, Jianqing (1996). Local Polynomial Modelling and Its Applications: From linear regression to nonlinear regression. Monographs on Statistics and Applied Probability. Chapman \& Hall/CRC. ISBN 978-0-412-98321-4.
%
\bibitem{Kermack} Kermack WO, McKendrick AG (1927) Contributions to the mathematical theory of epidemics. Proc R Soc A 115 :700–721.
%
\bibitem{Kourentzes:2019}
N. Kourentzes, Tutorial for the nnfor R package, Jan. 16, 2019, available on \url{https://kourentzes.com/forecasting/2019/01/16/tutorial-for-the-nnfor-r-package/}.
%
\bibitem{Newman:2010}
M.E.J. Newman, Networks An Introduction, Oxford University Press, 394. (2010).
%
\bibitem{bestfit}
S. Saxena, Underfitting vs. Overfitting (vs. Best Fitting) in Machine Learning, Feb. 7, 2020, Analytics Vidhya, , available on \url{https://www.analyticsvidhya.com}.
%
\bibitem{steven} Steven Sanche, Yen Ting Lin, Chonggang Xu, Ethan Romero-Severson, Nick Hengartner, and Ruian Ke. The novel coronavirus, 2019-ncov, is highly contagious and more infectious than initially estimated. medRxiv, 2020.
%
\bibitem{Vapnik}
Vapnik, V. The Nature of Statistical Learning Theory. Springer, New York, 1995.
%
\bibitem{msas} Minis\`ere de la sant\'e et de l'action sociale, Senegal, May 31, 2020, available on \url{http://www.sante.gouv.sn/}.
%
\bibitem{datahub}
COVID-19 Data Hub, available on \url{https://www.tableau.com/covid-19-coronavirus-}\\ \url{data-resources}.
%

%
\bibitem{prophet} Prophet: Automatic Forecasting Procedure, avalailable in \url{https://facebook.github.io/prophet/docs/} or  \url{https://github.com/facebook/prophet}.
%
\bibitem{python} Python Software Foundation. Python Language Reference, version 2.7.  Available at \url{http://www.python.org}.
%
\bibitem{CRAN}
CRAN package repository, available on \url{https://cran.r-project.org/web/packages/}.
\end{thebibliography}
\end{document}